\documentclass[lettersize,journal]{IEEEtran}

\usepackage{amsmath,amsfonts,amssymb,bm}
\usepackage{array,threeparttable,multirow,makecell}
\usepackage{graphicx,subcaption,caption}
\usepackage{float,stfloats,textcomp,url,verbatim,soul,cite}
\usepackage{fontawesome,pifont,tikz,setspace,xpatch}
\usepackage[export]{adjustbox}
\usepackage{lipsum}

\usepackage{algorithm}
\usepackage{algorithmic}



\renewcommand{\arraystretch}{1.1}

\soulregister{\cite}7 
\soulregister{\citep}7 
\soulregister{\citet}7 
\soulregister{\ref}7 
\soulregister{\pageref}7 
\begin{document}

\title{Byzantine Attacks in RIS-Enhanced Cooperative Spectrum Sensing: A Decision Fusion Perspective}
\author{Gaoyuan Zhang, Gaolei Song, Boyuan Li*, Zijian Li*, Baofeng Ji, Ruijuan Zheng, Guoqiang Zheng, Tony Q.S. Quek, \IEEEmembership{Fellow, IEEE}
\thanks{Gaoyuan Zhang, Gaolei Song, Baofeng Ji, Ruijuan Zheng and Guoqiang Zheng are with the School of Information Engineering, Henan University of Science and Technology, Luoyang 471023, China. They are also with the Science and Technology Innovation Center of Intelligent system, Longmen Laboratory, Luoyang 471000, China. Gaoyuan Zhang and Baofeng Ji are also with the Institute of Physics, Zhengzhou 451451, Henan Academy of Sciences. Email: gaoyuan.zhang@uestc.edu.cn.}
\thanks{Boyuan Li was with the School of Computer Science and
Artificial Intelligence, Zhengzhou University, Zhengzhou, China, 450001. Zijian Li is with Dalian Maritime University, College of Artificial Intelligence. Tony Q.S. Quek is with the Singapore University of Technology and Design,
Singapore 487372, and also with Yonsei Frontier Lab, Yonsei University,
Seoul 03722, South Korea. Email: tonyquek@sutd.edu.sg.}
\thanks{*Corresponding authors: Boyuan Li (Email: l202311841010602@gs.zzu .edu.cn) and Zijian Li (Email: lizj@dlmu.edu.cn)}
}

\markboth{Journal of \LaTeX\ Class Files,~Vol.~18, No.~9, September~2020}%
{How to Use the IEEEtran \LaTeX \ Templates}

\maketitle

\begin{abstract}
From the perspective of hard decision fusion, we investigate Byzantine attacks in Reconfigurable Intelligent Surface (RIS)-enhanced and decode-and-forward relay-assisted Cooperative Spectrum Sensing (CSS) for mobile Cognitive Radio Networks (CRNs) in this paper.  
Specially, a RIS-enhanced and decode-and-forward relay-assisted CSS configuration is first constructed under dynamic channel scenarios due to user mobility. Subsequently, the channel- and attack-aware hard decision fusion rules are developed, and the optimal channel-aware Byzantine attack strategies are then developed under both small-scale and large-scale attacking scenarios. 
The corresponding results depict that the optimal attack strategy does not require any a prior knowledge of the global instantaneous Channel State Information (ICSI) (e.g. false alarm probability and detection probability of all the secondary users), although perfect acquisition of ICSI is clearly always not affordable from the attacker perspective, which is further exacerbated by the RIS and decode-and-forward relays involved in CSS and the potential high mobility of secondary users that leads to fast fading channels. Furthermore, our counterintuitive results also indicate that, regardless of the attacker's awareness of the decision fusion rule, the optimal Byzantine attack can be achieved through a unifying framework, the explicit attack strategy may be not unique, and the attacking effectiveness is primarily determined by the fraction of the Byzantine nodes rather than the channel dynamics. That is, to make the channel-aware approach more practical, the challenge that the
heavy reliance on the global ICSI and decision fusion rule in obtaining the Byzantine attacks is successfully relaxed. Finally, we empirically validate our theoretical analysis through extensive simulations across a wide range of attacking scenarios.
\end{abstract}


\section{Introduction}
\subsection{Background} 
\IEEEPARstart{T}{HE} continuously exponential growth of lightweight IoT devices in 5G network is leading to increase in spectrum agility \cite{10663263,10947617,4}. Traditional static radio spectrum allocation policy is creating “artificial spectrum scarcity”, since Radio Frequency (RF) bands are either idle or highly underutilized most of the time and unlicensed users are not allowed to access \cite{10720658,10750460}. Cognitive Radio Networks (CRNs) is considered an enabling technology for efficient utilization of scarce RF spectrum to ameliorating the unavoidable spectrum scarcity by allowing opportunistic dynamic spectrum access to unlicensed secondary users (SUs) without causing any harmful interference to primary users (PUs) with the help of smart protocols and signal processing algorithms, such as spectrum sensing \cite{10398488,10955337,cai2024star}. 
Further, to prevail against various channel effects ( such as noise, shadowing, multipath fading etc.) and hidden terminal in real CRNs, Collaborative Spectrum Sensing (CSS) is imperative \cite{10064143,nasser2022intelligent,1}. CSS can incorporate the
spatial diversity to mitigate these problems, wherein different SUs work together with their independent observations and make a collaborative decision to distinguish between two hypotheses, e.g., the absence or presence of a spectrum hole \cite{10032816,10488747,2}.

An recent emerging technology named Reconfigurable Intelligent Surfaces (RIS) has added a new dimension to CSS design and demonstrated significant potential for augmenting CSS by reconfiguring the sensing and reporting channel characteristics even when CRNs suffer from a low signal-to-noise ratio (SNR) or strong interference \cite{9753781,10948318,10538434}. Unfortunately, this integratation is increasingly vulnerable to Spectrum Sensing Data Falsification (SSDF) or Byzantine attacks, wherein the compromised SUs can independently or collaboratively submit falsified sensing reports to blind the dedicated node called Fusion Center (FC) in infrastructure-based CRNs \cite{10354516,fu2023massive,10358357}. This is attributed to the fact that the unique and powerful channel reconfiguration of RIS can provide active degrees of freedom for attackers to exploit the capabilities of RIS-enhanced CSS to launch more complicated and unpredictable attacks with even greater damage \cite{luo2020attackers}. These intractable security threats are becoming more severe in large-scale CRNs, where massive sensing reports aggregation magnifies the influence of malicious reports \cite{7}.

To mitigate Byzantine attacks, a number of countermeasures for traditional CSS, including global decision-based defense \cite{kailkhura2015consensus}, reputation based approaches \cite{CHEN2024101921}, and abnormal
statistical-behavior detection based approaches \cite{10.1007/s11276-020-02398-w}, have been proposed.
For a detailed discussion and an extensive list of references of Byzantine attack defense, the author is referred to \cite{zhang2015byzantine}.
However, unlike those who were researching ways to use secure decision fusion or anomaly detection to prevail against Byzantine attacks in traditional CSS \cite{11006376}, we restrict our attention towards comprehensive assessment from an offensive perspective, and demonstrate how attackers can exploit the distributed decision making process to enable more potent attacks by sending false reports in RIS-enhanced CSS. 
Such adversary-oriented analysis can provide some profound insights of strategic attack-defense interplay for CSS in mobile CRNs given the available resources, which motivates our imperative study.

\subsection{Motivation and Contributions} 
To the best of our knowledge, there are still some shortcomings in current research on Byzantine attacks in CSS.
Firstly, for analysis simplification, classical studies address only the first level of uncertainty accounts for the PU phenomenon as observed by the SUs, the Byzantine attack strategy has pioneered the design under the premise that the sensing report transmission between the SUs and the FC is always reliable \cite{5624648}, and the second level of uncertainty typically due to the report channel between the SUs and the FC has not fully considered. However, the real CRNs are typically affected by receiver noise, channel fading, and interference. Further, the sensing report transmission is important from security point of view since it is one of the most vulnerable phases to attacks. Secondly, from the attacking effectiveness viewpoint, many researchers have extensively arrived a typical conclusion that the channel-aware design is attractive given the available resources considering its inherent adaptivity \cite{9609969}. 
In an intuitive sense, the channel-aware approach refers to the integration of
the knowledge of the global ICSI, such as the false alarm probability and detection probability for all the SUs, into the design of attacking algorithms. However, wireless communication is naturally uncertain and time-varying due to effects that are not always amenable to modeling, such as high mobility in 5G environment. Thus, perfect ICSI estimation or prediction is clearly complexity intensive and time-consuming, which is computationally challenging that attackers will meet and overcome operating in an adversarial environment and stringent resource constraints, most notably the energy constraint with compromised SUs operating on irreplaceable battery supply. Finally, the channel-aware attacking algorithm is highly sensitive to ICSI estimation and prediction \cite{9279294}. 
Most existing attack strategies under this theoretical framework have theoretical significance, but lack practical significance due to their very poor robustness when facing non-stationary communication channels. This paper aims to address the above mentioned issues and provide satisfactory solutions. The main contributions are summarized as follows:

\begin{itemize}
\item We propose a RIS-enhanced mobile CRN model subject to Byzantine attacks, where the report channel exhibit non-ideal characteristic. 
 
\item We develop the channel- and attack-aware hard decision fusion rules for the FC.
 
\item We analyze and optimize the channel-aware Byzantine attacks from the perspective of our developed hard decision fusion rules. 
 
\item We reveal that as long as the fraction of the Byzantine nodes remains unchanged, the optimal attack strategy remains consistent even if the adversary is in the absence of any side information about the global ICSI. 
 
\item We develop extensive simulations to verify our theoretical analysis across a wide range of attacking scenarios.
\end{itemize} 

It is worth pointed out that we pay all our attention towards distributed CSS under Byzantine attacks. The extension to hard decision fusion case in distributed wireless sensor networks (WSNs) under Byzantine attacks \cite{6482130} is straightforward but not pursued here. 
Moreover, the ensuing analysis is not tailored uniquely for distributed CSS, and thus the results of this paper may be easily extended to include any distributed communication configuration impaired by Byzantine attacks \cite{9411676}. 
Finally, most of the ensuing discussions can directly apply to multibit (soft) spectrum sensing output case \cite{10}, although it is not conceptually simple. Such analyses, however, will be under investigation in our future work.

\subsection{Paper Organization} 
To the best of our knowledge, this is the first work investigating the hard decision fusion to design Byzantine attacks in RIS-enhanced CSS for mobile CRNs. The remainder of this work takes the following structure. Section \ref{section2} introduces related research work on attack methods. Section \ref{section3} proposes a CSS network architecture. Section \ref{section4} develops the decision fusion rule for different scenarios. Section \ref{section5} analyzes the conditions that optimal attacks should satisfy when the optimal decision fusion rule is known. Section \ref{section6} explores the conditions that optimal attacks should satisfy when the suboptimal decision fusion rule is known. Section \ref{section7} provides some typical conclusions. Section \ref{section9} verifies the effectiveness of the theoretical analysis through simulation experiments. Section \ref{section10} provides the conclusions and prospects for future research direction. 

\section{Related Work}\label{section2}
We will discuss Byzantine attacks in Distributed WSNs, CSS, and Intelligent Cooperative Systems separately. Details of some related works are summarized in table \ref{table1}. 
\begin{table*}[!t]
  \small
  \begin{center}
    \caption{SUMMARY OF RELATED WORKS ON BYZANTINE ATTACKS}\label{table1}
    \scalebox{0.8}{
    \begin{tabular}{|c|c|c|c|cc|cc|c|}
    \hline
    \multirow{2}{*}{\textbf{\begin{tabular}[c]{@{}c@{}}{}\\{Works}\end{tabular}}} &\multirow{2}{*}{\textbf{\begin{tabular}[c]{@{}c@{}}{}\\{System Model}\end{tabular}}}                           & \multirow{2}{*}{\begin{tabular}[c]{@{}c@{}}\textbf{Performance}\\\textbf{Indicators}\end{tabular}}      & \multirow{2}{*}{\begin{tabular}[c]{@{}c@{}}\textbf{RIS}\\ \textbf{Enhancement}\end{tabular}}       & \multicolumn{2}{c|}{\textbf{\begin{tabular}[c]{@{}c@{}}{Antenna Number}\end{tabular}}}                      & \multicolumn{2}{c|}{\textbf{Required Priori Information}}                                                                                                                                      & \multirow{2}{*}{\begin{tabular}[c]{@{}c@{}}{}\\\textbf{Type of attack}\end{tabular}}          \\ \cline{5-8}
                                                                                  &                                                                                                               &                                                                                                         &                                                                                                    & \multicolumn{1}{c|}{\textbf{PU}}           & \multicolumn{1}{c|}{\textbf{SU}}                               & \multicolumn{1}{c|}{\textbf{${{P}_{{D}_{i}}}$/${{P}_{{F}_{i}}}$}}              & \multicolumn{1}{c|}{\begin{tabular}[c]{@{}c@{}}\textbf{True Sensing}\\\textbf{Information}\end{tabular} }    &         \\ \hline
    \cite{8}                                                                      & \begin{tabular}[c]{@{}c@{}}{Binary Hypothesis}\\ {Testing with}\\ {$M$-ary Detection}\end{tabular}            & \begin{tabular}[c]{@{}c@{}}{KLD}\end{tabular}                                                           & \multicolumn{1}{c|}{\ding{55}}                                                                     & \multicolumn{1}{c|}{Single}                & \multicolumn{1}{c|}{Single}                                    & \multicolumn{1}{c|}{\ding{52}}                                                  & \multicolumn{1}{c|}{\ding{52}}                                                                               & \multicolumn{1}{c|}{Static}                         \\ \hline
    \cite{9}                                                                      & \begin{tabular}[c]{@{}c@{}}{Binary Hypothesis}\\ {Testing with}\\ {Hard Detection}\end{tabular}               & \begin{tabular}[c]{@{}c@{}}{Chernoff Information}\end{tabular}                                          & \multicolumn{1}{c|}{\ding{55}}                                                                     & \multicolumn{1}{c|}{Single}                & \multicolumn{1}{c|}{Single}                                    & \multicolumn{1}{c|}{\ding{52}}                                                  & \multicolumn{1}{c|}{\ding{55}}                                                                               & \multicolumn{1}{c|}{Static }                         \\ \hline
    \cite{10}                                                                     & \begin{tabular}[c]{@{}c@{}}{Binary Hypothesis}\\ {Testing with}\\{Quantized Detection}\end{tabular}           & \begin{tabular}[c]{@{}c@{}}{Modified}\\{Deflection Coefficient}\end{tabular}                            & \multicolumn{1}{c|}{\ding{55}}                                                                    & \multicolumn{1}{c|}{Single}                 & \multicolumn{1}{c|}{Single}                                    & \multicolumn{1}{c|}{\ding{52}}                                                  & \multicolumn{1}{c|}{\ding{52}}                                                                               & \multicolumn{1}{c|}{Static }                         \\ \hline
    \cite{11}                                                                     & \begin{tabular}[c]{@{}c@{}}{Binary Hypothesis}\\ {Testing with}\\ {$M$-ary Detection}\end{tabular}            & \begin{tabular}[c]{@{}c@{}}{KLD}\end{tabular}                                                           & \multicolumn{1}{c|}{\ding{55}}                                                                    & \multicolumn{1}{c|}{Single}                 & \multicolumn{1}{c|}{Single}                                    & \multicolumn{1}{c|}{\ding{52}}                                                  & \multicolumn{1}{c|}{\ding{55}}                                                                               & \multicolumn{1}{c|}{Static }                         \\ \hline
    \cite{12}                                                                     & \begin{tabular}[c]{@{}c@{}}{$M$-ary Hypothesis}\\ {Testing with}\\ {Hard Detection}\end{tabular}              & \begin{tabular}[c]{@{}c@{}}{Minimum Fraction}\\{of Byzantine Nodes}\end{tabular}                        & \multicolumn{1}{c|}{\ding{55}}                                                                    & \multicolumn{1}{c|}{Single}                 & \multicolumn{1}{c|}{Single}                                    & \multicolumn{1}{c|}{\ding{52}}                                                  & \multicolumn{1}{c|}{\ding{52}}                                                                               & \multicolumn{1}{c|}{Static }                         \\ \hline
    \cite{13}                                                                     & \begin{tabular}[c]{@{}c@{}}{Binary Hypothesis}\\ {Testing with}\\ {Hard Detection}\end{tabular}               & \begin{tabular}[c]{@{}c@{}}{Global}\\{Detection performance}\end{tabular}                               & \multicolumn{1}{c|}{\ding{55}}                                                                    & \multicolumn{1}{c|}{Single}                 & \multicolumn{1}{c|}{Single}                                    & \multicolumn{1}{c|}{\ding{52}}                                                  & \multicolumn{1}{c|}{\ding{55}}                                                                               & \multicolumn{1}{c|}{Dynamic }                         \\ \hline
    \cite{14}                                                                     & \begin{tabular}[c]{@{}c@{}}{Binary Hypothesis}\\ {Testing with}\\ {Hard Detection}\end{tabular}               & \begin{tabular}[c]{@{}c@{}}{Malicious User}\\{Revenue}\end{tabular}              & \multicolumn{1}{c|}{\ding{55}}                                                                    & \multicolumn{1}{c|}{Single}                 & \multicolumn{1}{c|}{Single}                                    & \multicolumn{1}{c|}{\ding{52}}                                                  & \multicolumn{1}{c|}{\ding{55}}                                                                               & \multicolumn{1}{c|}{Dynamic}                         \\ \hline
    \cite{15}                                                                     & \begin{tabular}[c]{@{}c@{}}{Binary Hypothesis}\\ {Testing with}\\ {Hard Detection}\end{tabular}               & \begin{tabular}[c]{@{}c@{}}{MI}\end{tabular}                                    & \multicolumn{1}{c|}{\ding{55}}                                                                    & \multicolumn{1}{c|}{Single}                 & \multicolumn{1}{c|}{Single}                                    & \multicolumn{1}{c|}{\ding{55}}                                                  & \multicolumn{1}{c|}{\ding{55}}                                                                               & \multicolumn{1}{c|}{Dynamic}                         \\ \hline
    Ours                                                                          & \begin{tabular}[c]{@{}c@{}}{Binary Hypothesis}\\ {Testing with}\\ {Hard Detection}\end{tabular}               & \begin{tabular}[c]{@{}c@{}}{Reliability of Decision}\\{Fusion Statistics}\end{tabular}                  & \multicolumn{1}{c|}{\ding{52}}                                                                     & \multicolumn{1}{c|}{Single}                & \multicolumn{1}{c|}{$M$}                                         & \multicolumn{1}{c|}{\ding{55}}                                                  & \multicolumn{1}{c|}{\ding{55}}                                                                               & \multicolumn{1}{c|}{Dynamic}                         \\ \hline
    \end{tabular}}
  \end{center}
\end{table*}
\subsubsection{Byzantine Attacks in Distributed WSNs}
In the early studies on Byzantine attacks, it was commonly assumed that the true observation information was completely known. For example, under this assumption, \cite{8} formulated the problem as a Kullback-Leibler divergence (KLD) minimization and, through a “water-filling” procedure, derived the optimal attacking distributions for binary detection with quantized sensor observations. The study concluded that if the fraction of Byzantine nodes is less than half, the attackers cannot fully disrupt the system. However, although KLD is applicable for asymptotic analysis with a fixed number of sensors, it cannot directly provide accurate performance metrics. To overcome this limitation, \cite{9} analyzed the impact of Byzantine nodes on distributed Bayesian detection and employed Chernoff information as the performance metric, systematically investigating the asymptotic behavior of such systems. Unlike \cite{8}, which only obtained the optimal blinding strategy, \cite{9} further provided a closed-form expression for the optimal Byzantine attack.
To overcome the limitations of the aforementioned asymptotic analyses, Kailkhura et al. directly investigated the non-asymptotic scenario. Using the more direct probability of error as the performance metric, they derived the minimum fraction of attackers required to completely blind the system and provided closed-form solutions for the optimal attack strategies that maximize performance degradation when blinding is not achievable.
In a subsequent study, \cite{11} demonstrated, using the blinding probability, that uniform falsification can blind the FC, but only if the vast majority of nodes are compromised. To overcome this limitation, \cite{12} evaluated attack probabilities at the sensor level and, under the assumptions of symmetric falsification and that Byzantine attackers have access to local output statistics, derived an attack strategy capable of blinding a single sensor with the minimum compromise probability.

\subsubsection{Byzantine Attacks in CSS}
In the context of CSS, the complexity of sensing data and the critical need for accurate spectrum allocation make the system particularly vulnerable to sophisticated attacks. Consequently, research in this area has focused on more advanced attack models, such as those employing $M$-ary quantized data frameworks. For instance, \cite{10} conducted a more in-depth and systematic study of SSDF attacks under an $M$-ary quantized data framework based on the Neyman-Pearson criterion. It introduced a more sophisticated probabilistic SSDF attack model for CSS that the negative effect of defined probabilistic SSDF attack for the CSS with $M$-ary quantized data has been characterized, and the condition of the proposed SSDF attack model to make the FC completely incapable of inferring the status of PU has been derived. From a different perspective, \cite{8457252} proposes a robust evaluation benchmark from a FC perspective that integrates extended sensing with actual transmission feedback. Practical factors such as channel imperfections and inference errors are incorporated to focus on the detection of inconsistent sensors, providing a complementary perspective to the traditional analysis of blinding attacks. Similarly, \cite{9080068} examined attack strategies in the absence of any defense mechanisms, analyzing both attack strength and attack probability, and introduced a classical trust-based CSS algorithm to execute the attack strategies and evaluate the security of Byzantine attackers. These findings emphasize that a successful Byzantine strategy must carefully balance disruption and stealth, especially in systems employing trust or reputation mechanisms.
\subsubsection{Byzantine Attacks in Intelligent Cooperative Systems}
More recent research trends have shifted toward developing more realistic and adaptive attack models, particularly for intelligent cooperative systems. For example, \cite{13} proposed a dominant cooperative probability attack (DCPA) model, which aims to reduce detectability and enhance stealthiness. This DCPA model contains auxiliary cooperative attackers (ACAs) that launch attacks and a dominant attacker (DA) that determines the ACAs' attack probability intervals according to their respective credibility. By introducing dynamic probability cooperation (i.e., a non-fixed attack pattern), the model can guide the network's decision-making process. Compared with traditional static attack strategies, DCPA exhibits both higher stealth and greater destructiveness.
In another advanced application, \cite{10979375} proposed a probabilistic attack model for federated learning: given a target distribution, the model precisely derives the minimal amount of data contamination required to reduce model performance to the desired level. This work is the first to quantify the attack cost-performance degradation-stealthiness trade-off in a computable form. As a result, attackers can perform local, targeted, and hardly noticeable poisoning at extremely low cost. These developments represent a broader trend towards an adversarial model that is both strategically complex and attuned to the dynamics of the environment. This trend is exemplified by \cite{14}, which applied an evolutionary game theory framework allowing attackers to adaptively choose their flipping probability based on payoff optimization. Further advancing this line of inquiry, \cite{15} derived necessary and sufficient conditions for optimal attacks, showing that attackers can remain efficient and stealthy without CSI, thereby reducing energy consumption.
\subsubsection{Novelty of Our Work}
Compared with the previous works, the main novelty of this work is that the reliance on the global ICSI and decision fusion rules \cite{8, 11, 7134807} in obtaining the Byzantine attacks is fully relaxed, and robust Byzantine attacks can be successfully
implemented under stringent resource constraints over time-varying channels. Note here that, similar to the one considered here, a comprehensive coverage on Byzantine attacks in distributed WSNs can be found in \cite{15}. The distinction lies in that \cite{15} is presented from Mutual Information (MI) perspective without considering the RIS enhancement, whereas this work is developed based on decision fusion rules for RIS-enhanced CSS.

\section{System Model}\label{section3}
\begin{figure}[!t]
  \centering
       \includegraphics[width=0.7\linewidth]{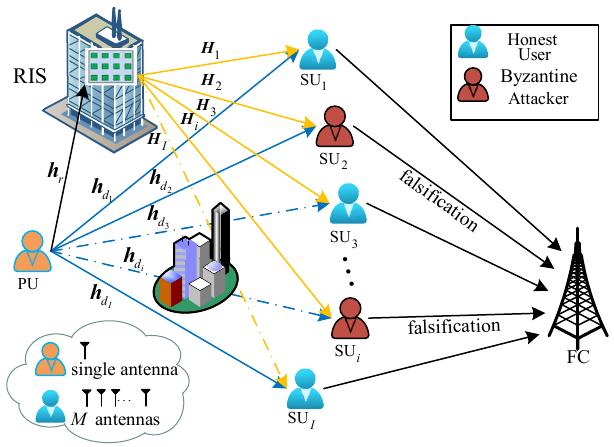}\\
  \caption{RIS Enhanced Mobile CRNs under Byzantine Attacks.}\label{L111}
\end{figure}

As depicted in Fig. \ref{L111}, we consider a more prevailing model, as not studied extensively in the CSS literature but arguably more relevant to engineering applications.
In our mobile CRN, there exists one PU with single antenna, $I$ SUs with $M$ antennas, one FC and an passive RIS with $N$ reflecting elements, where $N>I$. A RIS deployed within the cell is responsible for assisting SUs' spectrum sensing \cite{xie2024enhancing}. Normal SUs perform a detection approach for spectrum sensing, then report sensing results to the FC by employing multiple relay nodes. The FC then make global decision based on information gathered from local SUs to infer the absence or presence of a spectrum hole. Some SUs may be compromised by attackers and report falsified sensing results. 
In addition, the key variables used in this paper is listed in Tables \ref{table0}.

\begin{table*}[t]
  \footnotesize
  \centering
  \caption{DEFINITION OF THE KEY MATHEMATICAL NOTATIONS}
  \label{table0}
  \begin{tabular}{|c|l|c|l|}
  \hline
  \textbf{Symbols} & \textbf{Meaning} & \textbf{Symbols} & \textbf{Meaning} \\
  \hline
  $\mathbf{h}_{d_i}$ & The channel matrices from the PU to the $i$-th SU. & $\mathbf{h}_r$ & The channel matrices from the PU to the RIS. \\
  $\mathbf{H}_i$ & The channel matrices from the RIS to the $i$-th SU. & $x_i$ & Received signal at the $i$-th SU. \\
  $\bm{\Theta}$ & Adjustment phase matrix for RIS systems. & $p$ & The transmit power by the PU. \\
  $\mathbf{s}$ & The PU transmission symbol. & $\mathbf{n}_i$ & Gaussian noise at the $i$-th SU. \\
  $\mathbf{w}$ & The receive beamformer to equalize the received signal. & $f_s$ & The
  sampling frequency of the received signal. \\
  $\tau_s$ & The available sensing time. & $T$ & The signal sampling number ($T = \tau_s f_s$). \\
  $\lambda_i$ & The sensing threshold of the $i$-th SU. & $\gamma$ & The average SNR of the PU measured at the $i$-th SU.\\
  $P_{D_i}$ & Detection probability of the $i$-th SU. & $P_{F_i}$ &False alarm probability of the $i$-th SU.\\
   $\alpha$ & The intensity of the attack.& $Q$ & Q-function.\\
  $P_{1,0}$ & \begin{tabular}[l]{@{}l@{}}{The probability that an honest (Byzantine) node}\\{sends 1 to the FC when its actual spectrum sensing result is 0.}\end{tabular}  & $P_{0,1}$ & \begin{tabular}[l]{@{}l@{}}{The probability that an honest (Byzantine) node}\\{sends 0 to the FC when its actual spectrum sensing result is 1.}\end{tabular}\\
  \hline
  \end{tabular}
\end{table*}

\subsection{Sensing Channel Model}
The received signal for the $i$-th SU can be given by
\begin{align}
  \mathbf{x}_{i} = \left(\mathbf{h}_{d_i} + \mathbf{H}_{i} \bm{\Theta} \mathbf{h}_{r} \right) \sqrt{p} \mathbf{s} + \mathbf{n}_{i},
\end{align}
where ${\mathbf{h}_{d_i}\in {{\mathbb{C}}^{M\times 1}}}$, ${{\mathbf{h}}_{r}}\in {{\mathbb{C}}^{N\times 1}}$ and ${\mathbf{{H}}_{i}}\in {{\mathbb{C}}^{M\times N}}$ denote the channel matrices from the PU to the $i$-th SU, PU to the RIS, and RIS to the $i$-th SU, respectively, where $i=1,2,\ldots ,I$.  $p$ and $\mathbf{s}$ denote the PU's transmit power and information signal. The diagonal phase shift matrix for the RIS is defined as $\bm{\Theta} \triangleq \text{diag}\left[ e^{j\theta_1}, \cdots, e^{j\theta_N} \right]$, where $\quad \theta_n \in [0, 2\pi)$. $\mathbf{n}_{i} \sim \mathcal{CN}(0, \sigma^2{\mathbf{I}_{M}})$ represents AWGN at the $i$-th SU. 
The SU employs a receive beamformer ${\mathbf{w}\in {{\mathbb{C}}^{1\times M}}}$ to equalize the received signal, where \(\left\| \mathbf{w} \right\|^{2} = 1\), as 
\begin{align}
  {\mathbf{x}_{i}}={{\mathbf{w}}^{H}}{{\mathbf{x}_{i}}}={{\mathbf{w}}^{H}}\left( {{\mathbf{h}}_{d_i}}+{{\mathbf{H}_{i}}\Theta }{{\mathbf{h}}_{r}} \right)\sqrt{p}\mathbf{s}+{{\mathbf{w}}^{H}}\mathbf{n}_{i}.
\end{align}
Each SU has the ability to collect signal samples to determine whether the PU is active or not during the sensing interval. Therefore, we can rewrite the signal received by the $i$-th SU at the $t$-th sampling instant as
\begin{align}
x_i(t) = 
\begin{cases}
  {{\mathbf{w}}^{H}}\mathbf{n}_{i}(t), & H_0 \\
  {{\mathbf{w}}^{H}}[\left( \mathbf{h}_{d_i} + \mathbf{H}_{i} \bm{\Theta} \mathbf{h}_{r} \right) \sqrt{p} \mathbf{s}(t) + \mathbf{n}_{i}(t)], & H_1
\end{cases}
\end{align}
Energy detection is widely used in CRNs due to its simplicity and effectiveness \cite{wu2023sequential}. It does not require any a priori knowledge of PU's signal and strict phase
synchronization, so we only consider energy detection. Let $f_s$ be the sampling rate and $\tau_s$ the sensing time, so the number of samples is $T = \tau_s f_s$. Now the test statistic for energy detection is
\begin{equation}
  \resizebox{0.61\hsize}{!}{$\begin{aligned}
G(x) = \frac{1}{T} \sum_{t=1}^{T} |{x}_{i}(t)|^2, \quad t = 1,2,\dots,T,
\end{aligned}$}
\end{equation}
with the decision rule
\begin{equation}
  \resizebox{0.45\hsize}{!}{$\begin{aligned}
G(x) \mathop{\gtreqless}_{H_0}^{H_1} \lambda_i, \quad i = 1,2,\dots,I.
\end{aligned}$}
\end{equation}
where ${{\lambda }_{i}}$ is the pre-determined decision threshold for the $i$-th SU. For passive RIS, according to energy detection, the test statistic $G(x)$ can be written as follows
\begin{equation}
  \resizebox{0.91\hsize}{!}{$\begin{aligned}
    G(x)=
\begin{cases}
  \frac{1}{T} \sum\limits_{t=1}^{T}{{{\left| {{\mathbf{w}}^{H}}{\mathbf{n}_{i}}\left( t \right) \right|}^{2}},}&{{H}_{0}}  \\
  \frac{1}{T} \sum\limits_{t=1}^{T}{{{\left| {{\mathbf{w}}^{H}}[\left( {\mathbf{{h}}_{d_i}}+{{\mathbf{H}_{i}}\Theta }{{\mathbf{h}}_{r}} \right)\sqrt{p}\mathbf{s}(t)+{\mathbf{n}_{i}}\left( t \right)] \right|}^{2}}},&{{H}_{1}}  \\
  \end{cases}
\end{aligned}$}
\end{equation}
For larger values of $T$, the multidimensional Central Limit Theorem \cite{feller1991introduction} can be used to approximate $G(x)$, and modeled with a Gaussian distribution as follows
\begin{equation}
  \resizebox{0.91\hsize}{!}{$\begin{aligned}
    G(x)\sim 
\begin{cases}
  N\left( {{\sigma}^{2}},\frac{{{\sigma}^{4}}}{T} \right),&{{H}_{0}}  \\
  N\left[ p\left|{{\mathbf{w}}^{H}}\left( {\mathbf{{h}}_{d_i}}+{{\mathbf{H}_{i}}\Theta }{{\mathbf{h}}_{r}} \right) \right|+{{\sigma}^{2}},\frac{{{\sigma}^{4}}}{T}{{\left( 1+\gamma  \right)}^{2}} \right],&{{H}_{1}}  \\
\end{cases}
\end{aligned}$}
\end{equation}
where \(\gamma = \frac{p \left|{{\mathbf{w}}^{H}} (\mathbf{h}_{d_i} + {\mathbf{H}_{i}} \bm{\Theta} \mathbf{h}_r) \right|^2}{\sigma^2}\) is the instantaneous received SNR.
Since $G(x)$ follows a Gaussian distribution, thus the local false alarm probability ${P_{F_i}}$ and the local detection probability ${P_{D_i}}$ can be respectively expressed as
\begin{align}
  {P_{F_i}} &= Q\left[\left( \frac{\lambda_i}{\sigma^2} - 1 \right) \sqrt{T} \right],
\end{align}
and
\begin{align}
  {P_{D_i}} &= Q\left[\left( \frac{\lambda_i}{\sigma^2} - \gamma - 1 \right) \sqrt{\frac{T}{(1+\gamma)^2}} \right].
\end{align}
where $Q\left( x \right)=\frac{1}{\sqrt{2\pi }}\int_{x}^{\infty } \exp \left( -\frac{t^{2}}{2} \right) dt$ denotes the complementary cumulative distribution function of the standard Gaussian distribution.

\subsection{Byzantine Attacks Model}
We assume that each Byzantine attacker independently distort the local spectrum sensing reports of SUs to fulfill their malicious objectives. Specifically, we denote the probability that an honest (Byzantine) node sends 0 to the FC when its actual spectrum sensing result is 1 as $P_{0,1}$, while $P_{1,0}$ represents the probability that an honest (Byzantine) node sends 1 to the FC when its actual spectrum sensing result is 0. Denoting the manipulated data as $u_i$, we have
\begin{equation}\label{math10}
P(u_i = 0 \mid x_i = 1) \triangleq \pi_{0,1} = \alpha P_{0,1}
\end{equation}
and
 \begin{eqnarray}\label{math11}
P(u_i = 1 \mid x_i = 0) \triangleq \pi_{1,0} = \alpha P_{1,0},
\end{eqnarray}
where $\alpha$ denotes the fraction of the Byzantine nodes, i.e., the fraction of of SUs affected by the Byzantine attacks. Note here that we assume that each compromised node has the same
attacking probabilities $P_{0,1}$ and $P_{1,0}$. However, our analytical results can be extended to the case where compromised nodes have different attacking probabilities.

\subsection{Report Channel Model}
We assume that each relay node adopts the Decode-and-Forward (DF) strategy. In this context, each relay transmission process can be equivalently modeled as a Binary Channel (BC).
Let $\varepsilon_{i,j,0}$ and $\varepsilon_{i,j,1}$ denote the crossover probabilities of the $j$-th BC in the $i$-th path. Multiple BCs in serial configuration can be equivalent to a BC with crossover probabilities $\varepsilon_{i,0}$ and $\varepsilon_{i,1}$ \cite{10486923}. 
\begin{figure}[!h]
  \centering
       \includegraphics[width=0.9\linewidth]{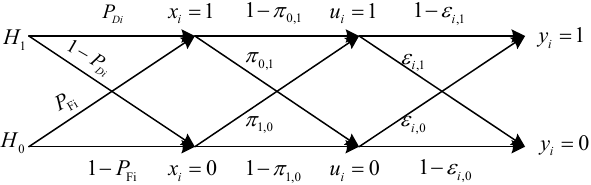}\\
       \captionsetup{skip=4pt}
  \caption{System Model with Byzantine Attacks. }\label{L1}
\end{figure}

It is worth pointed out that, based on above discussion, the equivalent transmission model can be represented in Fig. \ref{L1}. Note that we do not specify the sensing and report channel model or statistics of $\mathbf{h}_{d_i}$, ${{\mathbf{h}}_{r}}$, and ${\mathbf{H}_{i}}$. This need may arise due to the fact that we aim to provide a unifying framework in treating many of the unreliable communication channels in CSS of mobile CRNS. Also, we assume that the SUS are potential mobile.
Corresponding, the spectrum sensing channel and report channel may involve nonstationary fading, and the channel matrices $\mathbf{h}_{d_i}$, ${{\mathbf{h}}_{r}}$, and ${\mathbf{H}_{i}}$ are fast time-varying. This renders the previous approach assuming stationary fading channel statistics impractical and useless. However, the notations have been removed for simplicity. 

\section{Fusion Rules Under Byzantine Attacks}\label{section4}
We assume the FC is aware of Byzantine attacks and knows the attack parameters, including \(\alpha\), \(P_{1,0}\) and \(P_{0,1}\). However, the FC cannot identify specific compromised nodes and thus considers each SU to be Byzantine node with the certain probability $\alpha$. That is, a non-adversarial scenario is considered where Byzantine nodes do not actively confront the FC.

\subsection{Optimal Decision Fusion Rule}
Following from the ML criterion \cite{zhang2022multiple}, we can have that the decision fusion statistic for the FC can be given as
\begin{eqnarray}\label{math13}
  \begin{aligned}
\begin{array}{c}
  \Lambda \left( \mathbf{y} \right)=\ln \frac{P\left( \mathbf{y}|{{H}_{1}} \right)}{P\left( \mathbf{y}|{{H}_{0}} \right)}=\sum\limits_{i=1}^{I}{\ln \frac{P\left( {{y}_{i}}|{{H}_{1}} \right)}{P\left( {{y}_{i}}|{{H}_{0}} \right)}}=\sum\limits_{i=1}^{I}{\Lambda \left( {{y}_{i}} \right)},
\end{array}
\end{aligned}
\end{eqnarray}
where $\mathbf{y}=\left[ {{y}_{1}},{{y}_{2}},...,{{y}_{I}} \right]$ is the binary observation vector received at the FC from the $I$ branches after $(J-1)$ hops. The contribution of ${{y}_{i}}$ to the decision fusion statistic $\Lambda \left( \mathbf{y} \right)$ is denoted as $\Lambda \left( {{y}_{i}} \right)$, which can be expressed as (\ref{math14}) as shown at the bottom of the next page. 
\begin{figure*}[!b]
  \hrulefill
\begin{align}\label{math14}
  \Lambda \left( {{y}_{i}} \right) = \ln \frac{
   \begin{aligned}
        & {{P}_{{{D}_{i}}}}\left( 1-{{\pi }_{0,1}} \right)P\left( {{y}_{i}}|{{u}_{i}}=1 \right) + {{P}_{{{D}_{i}}}}{{\pi }_{0,1}}P\left( {{y}_{i}}|{{u}_{i}}=0 \right)  \\ 
        & +\left( 1-{{P}_{{{D}_{i}}}} \right){{\pi }_{1,0}}P\left( {{y}_{i}}|{{u}_{i}}=1 \right) + \left( 1-{{P}_{{{D}_{i}}}} \right)\left( 1-{{\pi }_{1,0}} \right)P\left( {{y}_{i}}|{{u}_{i}}=0 \right)
   \end{aligned}
      }{
   \begin{aligned}
        & {{P}_{{{F}_{i}}}}\left( 1-{{\pi }_{0,1}} \right)P\left( {{y}_{i}}|{{u}_{i}}=1 \right) + {{P}_{{{F}_{i}}}}{{\pi }_{0,1}}P\left( {{y}_{i}}|{{u}_{i}}=0 \right) \\ 
        & +\left( 1-{{P}_{{{F}_{i}}}} \right){{\pi }_{1,0}}P\left( {{y}_{i}}|{{u}_{i}}=1 \right) + \left( 1-{{P}_{{{F}_{i}}}} \right)\left( 1-{{\pi }_{1,0}} \right)P\left( {{y}_{i}}|{{u}_{i}}=0 \right)
      \end{aligned}
      }.
  \end{align}
\end{figure*}
When the output of the report channel ${{y}_{i}}=1$, $\Lambda \left( {{y}_{i}} \right)$ can be expressed as
\begin{eqnarray}\label{math15}
  \resizebox{0.9\hsize}{!}{$\begin{aligned}
  \Lambda \left( {{y}_{i}} \right)=\ln \frac{\begin{array}{l}
    {{P}_{{{D}_{i}}}}\left( 1-{{\pi }_{0,1}} \right)\left( 1-{{\varepsilon }_{i,1}} \right)+{{P}_{{{D}_{i}}}}{{\pi }_{0,1}}{{\varepsilon }_{i,0}} \\ 
    +\left( 1-{{P}_{{{D}_{i}}}} \right){{\pi }_{1,0}}\left( 1-{{\varepsilon }_{i,1}} \right) + \left( 1-{{P}_{{{D}_{i}}}} \right)\left( 1-{{\pi }_{1,0}} \right){{\varepsilon }_{i,0}}
  \end{array}}{\begin{array}{l}
    {{P}_{{{F}_{i}}}}\left( 1-{{\pi }_{0,1}} \right)\left( 1-{{\varepsilon }_{i,1}} \right)+{{P}_{{{F}_{i}}}}{{\pi }_{0,1}}{{\varepsilon }_{i,0}} \\ 
    +\left( 1-{{P}_{{{F}_{i}}}} \right){{\pi }_{1,0}}\left( 1-{{\varepsilon }_{i,1}} \right)+\left( 1-{{P}_{{{F}_{i}}}} \right)\left( 1-{{\pi }_{1,0}} \right){{\varepsilon }_{i,0}}
      \end{array}}.
\end{aligned}$}
\end{eqnarray}
When the output of the report channel ${{y}_{i}}=0$, $\Lambda \left( {{y}_{i}} \right)$ can be expressed as
\begin{eqnarray}\label{math16}
  \resizebox{0.9\hsize}{!}{$\begin{aligned}
  \Lambda \left( {{y}_{i}} \right)=\ln \frac{\begin{array}{l}
    {{P}_{{{D}_{i}}}}\left( 1-{{\pi }_{0,1}} \right){{\varepsilon }_{i,1}}+{{P}_{{{D}_{i}}}}{{\pi }_{0,1}}\left( 1-{{\varepsilon }_{i,0}} \right)\\
    +\left( 1-{{P}_{{{D}_{i}}}} \right){{\pi }_{1,0}}{{\varepsilon }_{i,1}}+\left( 1-{{P}_{{{D}_{i}}}} \right)\left( 1-{{\pi }_{1,0}} \right)\left( 1-{{\varepsilon }_{i,0}} \right)
  \end{array}}{\begin{array}{l}
    {{P}_{{F}_{i}}}\left( 1-{{\pi }_{0,1}} \right){{\varepsilon }_{i,1}}+{{P}_{{{F}_{i}}}}{{\pi }_{0,1}}\left( 1-{{\varepsilon }_{i,0}} \right)\\
    +\left( 1-{{P}_{{{F}_{i}}}} \right){{\pi }_{1,0}}{{\varepsilon }_{i,1}}+\left( 1-{{P}_{{{F}_{i}}}} \right)\left( 1-{{\pi }_{1,0}} \right)\left( 1-{{\varepsilon }_{i,0}} \right)
      \end{array}}.
\end{aligned}$}
\end{eqnarray}

Bringing (\ref{math15}) and (\ref{math16}) into (\ref{math14}) yields the optimal decision fusion statistic $\Lambda \left( \mathbf{y} \right)$, which can be expressed as (\ref{math17}), shown at the bottom of the next page.
\begin{figure*}[!b]
  \begin{align}\label{math17}
    \Lambda \left( \mathbf{y} \right)
   &{\rm{ = }}\sum\limits_{i:{{y}_{i}}=1} {\ln \frac{{{P}_{{{D}_{i}}}}\left( 1-{{\pi }_{0,1}} \right)\left( 1-{{\varepsilon }_{i,1}} \right)+{{P}_{{{D}_{i}}}}{{\pi }_{0,1}}{{\varepsilon }_{i,0}}+\left( 1-{{P}_{{{D}_{i}}}} \right){{\pi }_{1,0}}\left( 1-{{\varepsilon }_{i,1}} \right)+\left( 1-{{P}_{{{D}_{i}}}} \right)\left( 1-{{\pi }_{1,0}} \right){{\varepsilon }_{i,0}}}{{{P}_{{{F}_{i}}}}\left( 1-{{\pi }_{0,1}} \right)\left( 1-{{\varepsilon }_{i,1}} \right)+{{P}_{{{F}_{i}}}}{{\pi }_{0,1}}{{\varepsilon }_{i,0}}+\left( 1-{{P}_{{{F}_{i}}}} \right){{\pi }_{1,0}}\left( 1-{{\varepsilon }_{i,1}} \right)+\left( 1-{{P}_{{{F}_{i}}}} \right)\left( 1-{{\pi }_{1,0}} \right){{\varepsilon }_{i,0}}}} \notag \\ 
   &\quad{\rm{ + }}\sum\limits_{i:{{y}_{i}}=0} {\ln \frac{{{P}_{{{D}_{i}}}}\left( 1-{{\pi }_{0,1}} \right){{\varepsilon }_{i,1}}+{{P}_{{{D}_{i}}}}{{\pi }_{0,1}}\left( 1-{{\varepsilon }_{i,0}} \right)+\left( 1-{{P}_{{{D}_{i}}}} \right){{\pi }_{1,0}}{{\varepsilon }_{i,1}}+\left( 1-{{P}_{{{D}_{i}}}} \right)\left( 1-{{\pi }_{1,0}} \right)\left( 1-{{\varepsilon }_{i,0}} \right)}{{{P}_{{{F}_{i}}}}\left( 1-{{\pi }_{0,1}} \right){{\varepsilon }_{i,1}}+{{P}_{{{F}_{i}}}}{{\pi }_{0,1}}\left( 1-{{\varepsilon }_{i,0}} \right)+\left( 1-{{P}_{{{F}_{i}}}} \right){{\pi }_{1,0}}{{\varepsilon }_{i,1}}+\left( 1-{{P}_{{{F}_{i}}}} \right)\left( 1-{{\pi }_{1,0}} \right)\left( 1-{{\varepsilon }_{i,0}} \right)}}.
  \end{align}
\end{figure*}

\subsection{Suboptimal Decision Fusion Rules under Ideal Local Sensing Channels}
Under the ideal condition of the local sensing channel, we have $P_{D_i} = 1$ and $P_{F_i} = 0$, and (\ref{math15}) can be rewritten as
\begin{eqnarray}\label{math18}
  \begin{aligned}
    \Lambda \left( {{y}_{i}} \right)=\ln \frac{\left( 1-{{\varepsilon }_{i,1}} \right)-{{\pi }_{0,1}}\left( 1-{{\varepsilon }_{i,0}}-{{\varepsilon }_{i,1}} \right)}{{{\varepsilon }_{i,0}}+{{\pi }_{1,0}}\left( 1-{{\varepsilon }_{i,0}}-{{\varepsilon }_{i,1}} \right)}.
\end{aligned}
\end{eqnarray}
Similarly, (\ref{math16}) can be rewritten as
\begin{eqnarray}\label{math19}
  \begin{aligned}
\Lambda \left( {{y}_{i}} \right)=\ln \frac{{{\varepsilon }_{i,1}}+{{\pi }_{0,1}}\left( 1-{{\varepsilon }_{i,0}}-{{\varepsilon }_{i,1}} \right)}{\left( 1-{{\varepsilon }_{i,0}} \right)-{{\pi }_{1,0}}\left( 1-{{\varepsilon }_{i,0}}-{{\varepsilon }_{i,1}} \right)}.
\end{aligned}
\end{eqnarray}

\subsection{Suboptimal Fusion Rules when the SNR for the Multihop Relay Network is High}
At this point, there is ${{\varepsilon}_{i,0}}\to 0$ and ${{\varepsilon}_{i,1}}\to 0$, then (\ref{math15}) can be rewritten as
\begin{eqnarray}\label{math20}
  \begin{aligned}
\Lambda \left( {{y}_{i}} \right)=\ln \frac{{{P}_{{{D}_{i}}}}\left( 1-{{\pi }_{0,1}} \right)+\left( 1-{{P}_{{{D}_{i}}}} \right){{\pi }_{1,0}}}{{{P}_{{{F}_{i}}}}\left( 1-{{\pi }_{0,1}} \right)+\left( 1-{{P}_{{{F}_{i}}}} \right){{\pi }_{1,0}}},
\end{aligned}
\end{eqnarray}
and (\ref{math16}) can be rewritten as
\begin{eqnarray}\label{math21}
  \begin{aligned}
    \Lambda \left( {{y}_{i}} \right)=\ln \frac{{{P}_{{{D}_{i}}}}{{\pi }_{0,1}}+\left( 1-{{P}_{{{D}_{i}}}} \right)\left( 1-{{\pi }_{1,0}} \right)}{{{P}_{{{F}_{i}}}}{{\pi }_{0,1}}+\left( 1-{{P}_{{{F}_{i}}}} \right)\left( 1-{{\pi }_{1,0}} \right)}.
  \end{aligned}
\end{eqnarray}

\subsection{Suboptimal Decision Fusion Rules when the SNR for the Multihop Relay Network is Low}
When the SNR for each relay channel is low, each relay transmission process can be approximately modeled as a Binary Symmetric Channel (BSC) according to the fact that ${{\varepsilon }_{i,j,0}} \to 0.5$ and ${{\varepsilon }_{i,j,1}} \to 0.5$. In this context, we have that ${{\varepsilon }_{i,0}}={{\varepsilon }_{i,1}} \approx 0.5$. Let ${{\delta }_{i,j}}=\log \frac{1-{{\varepsilon }_{i,1}}}{{{\varepsilon }_{i,0}}}$, following from the fact that \cite{ZHANG2022167}
\begin{eqnarray}\label{math22}
  \begin{aligned}
  \frac{1-{{\varepsilon }_{i,1}}}{{{\varepsilon }_{i,0}}}&=\frac{\frac{1}{2}\left[ 1+\prod{_{j=1}^{J}}\left( 1-2{{\varepsilon }_{i,j,1}} \right) \right]}{1-\frac{1}{2}\left[ 1+\prod{_{j=1}^{J}}\left( 1-2{{\varepsilon }_{i,j,0}} \right) \right]}\\
 &\approx \exp \left[ {{\min }_{1\le j\le J}}\left( {{\delta }_{i,j}} \right) \right],
\end{aligned}
\end{eqnarray}
we can directly rewrite (\ref{math15}) as (\ref{math23}), which is shown at the bottom of the next page.
\begin{figure*}[!b]
  \hrulefill
  \begin{align}\label{math23}
  \Lambda \left( {{y}_{i}} \right)&=\ln \frac{{{P}_{{{D}_{i}}}}\left( 1-{{\pi }_{0,1}} \right)\left( 1-{{\varepsilon }_{i,1}} \right)+{{P}_{{{D}_{i}}}}{{\pi }_{0,1}}{{\varepsilon }_{i,0}}+\left( 1-{{P}_{{{D}_{i}}}} \right){{\pi }_{1,0}}\left( 1-{{\varepsilon }_{i,1}} \right)+\left( 1-{{P}_{{{D}_{i}}}} \right)\left( 1-{{\pi }_{1,0}} \right){{\varepsilon }_{i,0}}}{{{P}_{{{F}_{i}}}}\left( 1-{{\pi }_{0,1}} \right)\left( 1-{{\varepsilon }_{i,1}} \right)+{{P}_{{{F}_{i}}}}{{\pi }_{0,1}}{{\varepsilon }_{i,0}}+\left( 1-{{P}_{{{F}_{i}}}} \right){{\pi }_{1,0}}\left( 1-{{\varepsilon }_{i,1}} \right)+\left( 1-{{P}_{{{F}_{i}}}} \right)\left( 1-{{\pi }_{1,0}} \right){{\varepsilon }_{i,0}}} \notag \\ 
  &\approx \ln \frac{\left[ {{P}_{{{D}_{i}}}}{{\pi }_{0,1}}+\left( 1-{{P}_{{{D}_{i}}}} \right)\left( 1-{{\pi }_{1,0}} \right) \right]\frac{{{\varepsilon }_{i,0}}}{{{\varepsilon }_{i,1}}}+\left[ {{P}_{{{D}_{i}}}}\left( 1-{{\pi }_{0,1}} \right)+\left( 1-{{P}_{{{D}_{i}}}} \right){{\pi }_{1,0}} \right]\exp \left[ {{\min }_{1\le j\le J}}\left( {{\delta }_{i,j}} \right) \right]}{\left[ {{P}_{{{F}_{i}}}}{{\pi }_{0,1}}+\left( 1-{{P}_{{{F}_{i}}}} \right)\left( 1-{{\pi }_{1,0}} \right) \right]\frac{{{\varepsilon }_{i,0}}}{{{\varepsilon }_{i,1}}}+\left[ {{P}_{{{F}_{i}}}}\left( 1-{{\pi }_{0,1}} \right)+\left( 1-{{P}_{{{F}_{i}}}} \right){{\pi }_{1,0}} \right]\exp \left[ {{\min }_{1\le j\le J}}\left( {{\delta }_{i,j}} \right) \right]}.
\end{align}
\end{figure*}
Under low channel SNR conditions, the exponential term in (\ref{math23}) can be accurately approximated by its first-order Taylor expansion without introducing unnecessary errors \cite{10552422}, we can immediately simplify (\ref{math23}) as:
\begin{eqnarray}\label{math24}
  \resizebox{0.9\hsize}{!}{$\begin{aligned}
  \Lambda \left( {{y}_{i}} \right)\approx \ln \frac{1+\left[ {{P}_{{{D}_{i}}}}\left( 1-{{\pi }_{0,1}} \right)+\left( 1-{{P}_{{{D}_{i}}}} \right){{\pi }_{1,0}} \right]{{\min }_{1\le j\le J}}\left( {{\delta }_{i,j}} \right)}{1+\left[ {{P}_{{{F}_{i}}}}\left( 1-{{\pi }_{0,1}} \right)+\left( 1-{{P}_{{{F}_{i}}}} \right){{\pi }_{1,0}} \right]{{\min }_{1\le j\le J}}\left( {{\delta }_{i,j}} \right)}.
\end{aligned}$}
\end{eqnarray}
Using the fact that, $x\to 0$, $\log \left( 1+x \right)\to x$, (\ref{math24}) can be further simplified as
\begin{eqnarray}\label{math25}
  \begin{aligned}
   \Lambda \left( {{y}_{i}} \right)
 & \approx \left( 1-{{\pi }_{0,1}}-{{\pi }_{1,0}} \right)\left( {{P}_{{{D}_{i}}}}-{{P}_{{{F}_{i}}}} \right)\underset{1\le j\le J}{\mathop{\min }}\,\left( {{\delta }_{i,j}} \right).  
  \end{aligned}
\end{eqnarray}
Similarly, when ${{y}_{i}}=0$, we can easily have that
\begin{eqnarray}\label{math26}
  \begin{aligned}
\Lambda \left( {{y}_{i}} \right)\approx -\left( 1-{{\pi }_{0,1}}-{{\pi }_{1,0}} \right)\left( {{P}_{{{D}_{i}}}}-{{P}_{{{F}_{i}}}} \right)\underset{1\le j\le J}{\mathop{\min }}\,\left( {{\delta }_{i,j}} \right).
\end{aligned}
\end{eqnarray}

\section{Optimal Byzantine Attacks when the Optimal Decision Fusion Rule is Known}\label{section5}
\subsection{Optimal Small-Scale Byzantine Attack}
For the convenience of analysis, we start by respectively modifying the optimal LLR-based decision fusion statistics given in (\ref{math15}) and (\ref{math16}) as
\begin{eqnarray}\label{math31}
  \resizebox{0.9\hsize}{!}{$\begin{aligned}
  &\Lambda \left( {{y}_{i}} \right)=\ln \frac{\begin{array}{l}
    {{P}_{{{D}_{i}}}}\left( 1-{{\varepsilon }_{i,1}} \right)+\left( 1-{{P}_{{{D}_{i}}}} \right){{\varepsilon }_{i,0}}\\
    -\left[ {{P}_{{{D}_{i}}}}{{\pi }_{0,1}}-\left( 1-{{P}_{{{D}_{i}}}} \right){{\pi }_{1,0}} \right]\left( 1-{{\varepsilon }_{i,0}}-{{\varepsilon }_{i,1}} \right)
  \end{array}}{\begin{array}{l}
    {{P}_{{{F}_{i}}}}\left( 1-{{\varepsilon }_{i,1}} \right)+\left( 1-{{P}_{{{F}_{i}}}} \right){{\varepsilon }_{i,0}}\\
    -\left[ {{P}_{{{F}_{i}}}}{{\pi }_{0,1}}-\left( 1-{{P}_{{{F}_{i}}}} \right){{\pi }_{1,0}} \right]\left( 1-{{\varepsilon }_{i,0}}-{{\varepsilon }_{i,1}} \right)
      \end{array}},
\end{aligned}$}
\end{eqnarray}
and
\begin{eqnarray}\label{math32}
  \resizebox{0.9\hsize}{!}{$\begin{aligned}
\Lambda \left( {{y}_{i}} \right)=\ln \frac{\begin{array}{l}
  {{P}_{{{D}_{i}}}}{{\varepsilon }_{i,1}}+\left( 1-{{P}_{{{D}_{i}}}} \right)\left( 1-{{\varepsilon }_{i,0}} \right)\\
  +\left[ {{P}_{{{D}_{i}}}}{{\pi }_{0,1}}-\left( 1-{{P}_{{{D}_{i}}}} \right){{\pi }_{1,0}} \right]\left( 1-{{\varepsilon }_{i,0}}-{{\varepsilon }_{i,1}} \right)
\end{array}}{\begin{array}{l}
  {{P}_{{{F}_{i}}}}{{\varepsilon }_{i,1}}+\left( 1-{{P}_{{{F}_{i}}}} \right)\left( 1-{{\varepsilon }_{i,0}} \right)\\
  +\left[ {{P}_{{{F}_{i}}}}{{\pi }_{0,1}}-\left( 1-{{P}_{{{F}_{i}}}} \right){{\pi }_{1,0}} \right]\left( 1-{{\varepsilon }_{i,0}}-{{\varepsilon }_{i,1}} \right)
\end{array}}.
\end{aligned}$}
\end{eqnarray}

We first pay our attention towards (\ref{math31}). Here, the fraction of Byzantine nodes satisfies $\alpha \le 0.5$, thus for any given ${{P}_{{{D}_{i}}}}$, ${{P}_{{F}_{{i}}}}$, ${{\varepsilon }_{i,0}}$ and ${{\varepsilon }_{i,1}}$, provided that ${{{\pi }_{0,1}}}={\pi }_{1,0}$, then (\ref{math31}) can be  reduced to
\begin{eqnarray}\label{math33}
  \begin{aligned}
\Lambda \left( {{y}_{i}} \right)=\ln \frac{\begin{array}{l}
  {{P}_{{{D}_{i}}}}\left( 1-{{\varepsilon }_{i,1}} \right)+\left( 1-{{P}_{{{D}_{i}}}} \right){{\varepsilon }_{i,0}}\\
  -{{\pi }_{0,1}}\left( 2{{P}_{{{D}_{i}}}}-1 \right)\left( 1-{{\varepsilon }_{i,0}}-{{\varepsilon }_{i,1}} \right)
\end{array}}{\begin{array}{l}
  {{P}_{{{F}_{i}}}}\left( 1-{{\varepsilon }_{i,1}} \right)+\left( 1-{{P}_{{{F}_{i}}}} \right){{\varepsilon }_{i,0}}\\
  +{{\pi }_{0,1}}\left( 1-2{{P}_{{{F}_{i}}}} \right)\left( 1-{{\varepsilon }_{i,0}}-{{\varepsilon }_{i,1}} \right)
\end{array}}.
\end{aligned}
\end{eqnarray}
Further, it is evident from (\ref{math33}) that \(2P_{D_i} - 1 > 0\) and \(1 - 2P_{F_i} > 0\). If both \(\pi_{0,1}\) and \(\pi_{1,0}\) reach their maximum value \(\alpha\), 
i.e., \(P_{0,1} = P_{1,0} = 1\), the numerator of (\ref{math33}) reaches its minimum while the denominator attains its maximum. 
Considering that the numerator is larger than the denominator, then the magnitude of $\Lambda(y_i)$ achieves the minimum at this point, and thus results in the lowest reliability of $\Lambda(y_i)$.

Similarly, substituting ${{\pi }_{0,1}}={{\pi }_{1,0}}$ into (\ref{math32}) yields
\begin{eqnarray}\label{math34}
  \resizebox{0.75\hsize}{!}{$\begin{aligned}
\Lambda \left( {{y}_{i}} \right)=\ln \frac{\begin{array}{l}
  {{P}_{{{D}_{i}}}}{{\varepsilon }_{i,1}}+\left( 1-{{P}_{{{D}_{i}}}} \right)\left( 1-{{\varepsilon }_{i,0}} \right)\\
  +{{\pi }_{0,1}}\left( 2{{P}_{{{D}_{i}}}}-1 \right)\left( 1-{{\varepsilon }_{i,0}}-{{\varepsilon }_{i,1}} \right)
\end{array}}{\begin{array}{l}
  {{P}_{{{F}_{i}}}}{{\varepsilon }_{i,1}}+\left( 1-{{P}_{{{F}_{i}}}} \right)\left( 1-{{\varepsilon }_{i,0}} \right)\\
  -{{\pi }_{0,1}}\left( 1-2{{P}_{{{F}_{i}}}} \right)\left( 1-{{\varepsilon }_{i,0}}-{{\varepsilon }_{i,1}} \right)
\end{array}}.
\end{aligned}$}
\end{eqnarray}
If both \(\pi_{0,1}\) and \(\pi_{1,0}\) reach their maximum value \(\alpha\), 
i.e., \(P_{0,1} = P_{1,0} = 1\), the numerator of (\ref{math34}) reaches its maximum while the denominator attains its minimum. 
Considering that the numerator is smaller than the denominator, then the magnitude of $\Lambda(y_i)$ achieves the minimum, and also results in the lowest reliability of $\Lambda(y_i)$.

Finally, it is crucial to note from (\ref{math33}) and (\ref{math34}) that the fraction of Byzantine nodes $\alpha$, the probabilities ${P}_{0,1}$ and ${{P}_{1,0}}$, do not independently affect the decision fusion statistic $\Lambda \left( {{y}_{i}} \right)$. In fact, they are intertwined with the detection probability ${{P}_{{D}_{i}}}$, the false alarm probability ${{P}_{{F}_{i}}}$, and the report channel ICSI ${{\varepsilon }_{i,0}}$ and ${{\varepsilon }_{i,1}}$, and affect the magnitude of $\Lambda(y_i)$ in an interdependent manner.
For example, in the case of the Always-Yes (AY) attack, we have ${{P}_{0,1}}=0$ and ${{P}_{1,0}}=1$. Correspondingly, we have ${{\pi }_{0,1}}=0$ and ${{\pi }_{1,0}}=\alpha$, leading to
\begin{eqnarray}\label{math35}
  \resizebox{0.9\hsize}{!}{$\begin{aligned}
\left\{
  \begin{aligned}
   &{{P}_{{{D}_{i}}}}{{\pi }_{0,1}}-\left( 1-{{P}_{{{D}_{i}}}} \right){{\pi }_{1,0}}=\left( {{P}_{{{D}_{i}}}}-1 \right)\alpha <0\\
   &{{P}_{{{F}_{i}}}}{{\pi }_{0,1}}-\left( 1-{{P}_{{{F}_{i}}}} \right){{\pi }_{1,0}}=\left( {{P}_{{{F}_{i}}}}-1 \right)\alpha <0\\
   &\left| {{P}_{{{D}_{i}}}}{{\pi }_{0,1}}-\left( 1-{{P}_{{{D}_{i}}}} \right){{\pi }_{1,0}} \right|<\left| {{P}_{{{F}_{i}}}}{{\pi }_{0,1}}-\left( 1-{{P}_{{{F}_{i}}}} \right){{\pi }_{1,0}} \right|.  \\
   \end{aligned}
\right.
\end{aligned}$}
\end{eqnarray}

Following the result in (\ref{math35}), analysis of (\ref{math31}) and (\ref{math32}) shows that, compared to the no-attack case ($P_{0,1}=0$, $P_{1,0}=0$), the increment of the numerator in (\ref{math31}) is less than that of the denominator, while in (\ref{math32}) the decreasment of the numerator is less than that of the denominator, and the numerator and denominator merge and gradually become identical. Consequently, both the magnitude and reliability of $\Lambda(y_i)$ are degraded.
Similarly, for the  Always-No (AN) attack, ${{P}_{0,1}}=1$ and ${{P}_{1,0}}=0$, there are ${{\pi }_{0,1}}=\alpha$ and ${{\pi }_{1,0}}=0$, leading to
\begin{eqnarray}\label{math36}
  \resizebox{0.9\hsize}{!}{$\begin{aligned}
\left\{ 
\begin{aligned}
& P_{D_i} \pi_{0,1} - (1 - P_{D_i}) \pi_{1,0} = \alpha P_{D_i} > 0, \\
& P_{F_i} \pi_{0,1} - (1 - P_{F_i}) \pi_{1,0} = \alpha P_{F_i} > 0, \\
& \left| P_{D_i} \pi_{0,1} - (1 - P_{D_i}) \pi_{1,0} \right| > \left| P_{F_i} \pi_{0,1} - (1 - P_{F_i}) \pi_{1,0} \right|.
\end{aligned}
\right.
\end{aligned}$}
\end{eqnarray}
The same observation can be easily drawn as we depict for AY attack.
Following the results reported in (\ref{math33}) and (\ref{math34}), we can immediately achieve an argument that for small-scale attacks, the optimal strategy is the \textbf{Always-False (AF) attack}. This is achieved by setting the flip probabilities to their maximum, i.e., $\pi_{0,1} = \pi_{1,0} = \alpha$ (corresponding to $P_{0,1} = P_{1,0} = 1$).

\subsection{Optimal Large-Scale Byzantine Attack}
When the fraction of Byzantine nodes satisfies $\alpha > 0.5$, for any given values of $P_{D_i}$, $P_{F_i}$, ${{\varepsilon }_{i,0}}$ and ${{\varepsilon }_{i,1}}$, substituting $\pi_{0,1} + \pi_{1,0} = 1$ into (\ref{math31}) and (\ref{math32}) yields the following results. When the output of the report channel ${{y}_{i}}=1$
\begin{eqnarray}\label{math37}
  \begin{aligned}
\Lambda \left( {{y}_{i}} \right)=\ln \frac{{{\varepsilon }_{i,0}}+{{\pi }_{0,1}}\left( 1-{{\varepsilon }_{i,0}}-{{\varepsilon }_{i,1}} \right)}{{{\varepsilon }_{i,0}}+{{\pi }_{0,1}}\left( 1-{{\varepsilon }_{i,0}}-{{\varepsilon }_{i,1}} \right)}=0,
\end{aligned}
\end{eqnarray}
when the output of the report channel ${{y}_{i}}=0$
\begin{eqnarray}\label{math38}
  \begin{aligned}
\Lambda \left( {{y}_{i}} \right)=\ln \frac{1-{{\varepsilon }_{i,0}}-{{\pi }_{0,1}}\left( 1-{{\varepsilon }_{i,0}}-{{\varepsilon }_{i,1}} \right)}{1-{{\varepsilon }_{i,0}}-{{\pi }_{0,1}}\left( 1-{{\varepsilon }_{i,0}}-{{\varepsilon }_{i,1}} \right)}=0.
\end{aligned}
\end{eqnarray}

Following the results reported in (\ref{math37}) and (\ref{math38}), we have that when $\alpha > 0.5$, for any $P_{D_i}$, $P_{F_i}$, $\varepsilon_{i,0}$ and $\varepsilon_{i,1}$, any attack strategy satisfying $\pi_{0,1} + \pi_{1,0} = 1$ forces $\Lambda(y_i)$ to 0, making it completely unreliable. In this context, the optimal attack is the \textbf{Random (RD) attack}, which satisfies $\pi_{0,1} + \pi_{1,0} = 1$. The explicit attack strategy is implicitly not unique.

\section{Optimal Byzantine Attacks when the suboptimal Decision fusion rule is known}\label{section6}
\subsection{Optimal Small-Scale Byzantine Attack under Ideal Local Sensing Channels}
For the suboptimal fusion rule under ideal sensing channel, when the fraction of Byzantine attack satisfies $\alpha \le 0.5$, for any given ${{P}_{{{D}_{i}}}}$, ${{P}_{{F}_{{i}}}}$, ${{\varepsilon }_{i,0}}$ and ${{\varepsilon }_{i,1}}$, provided that ${{\pi }_{0,1}}={{\pi }_{1,0}}$, (\ref{math18}) can be changed to
\begin{eqnarray}\label{math47}
  \begin{aligned}
\Lambda \left( {{y}_{i}} \right)=\ln \frac{1-\left[ {{\varepsilon }_{i,1}}+{{\pi }_{0,1}}\left( 1-{{\varepsilon }_{i,0}}-{{\varepsilon }_{i,1}} \right) \right]}{{{\varepsilon }_{i,0}}+{{\pi }_{0,1}}\left( 1-{{\varepsilon }_{i,0}}-{{\varepsilon }_{i,1}} \right)},
\end{aligned}
\end{eqnarray}
and (\ref{math19}) can be changed to
\begin{eqnarray}\label{math48}
  \begin{aligned}
\Lambda \left( {{y}_{i}} \right)=\ln \frac{{{\varepsilon}_{i,1}}+{{\pi }_{0,1}}\left( 1-{{\varepsilon }_{i,0}}-{{\varepsilon }_{i,1}} \right)}{1-\left[ {{\varepsilon }_{i,0}}+{{\pi }_{0,1}}\left( 1-{{\varepsilon }_{i,0}}-{{\varepsilon }_{i,1}} \right) \right]}.
\end{aligned}
\end{eqnarray}

The derivative of (\ref{math47}) with respect to ${{\pi }_{0,1}}$ yields
\begin{eqnarray}\label{math49}
  \resizebox{0.9\hsize}{!}{$\begin{aligned}
\frac{\partial \Lambda \left( {{y}_{i}} \right)}{\partial {{\pi }_{0,1}}}=\frac{-\left( 1-{{\varepsilon }_{i,0}}-{{\varepsilon }_{i,1}} \right)\left( 1+{{\varepsilon }_{i,0}}-{{\varepsilon }_{i,1}} \right)}{\left[ {{\varepsilon }_{i,0}}+{{\pi }_{0,1}}\left( 1-{{\varepsilon }_{i,0}}-{{\varepsilon }_{i,1}} \right) \right]\left\{ 1-\left[ {{\varepsilon }_{i,1}}+{{\pi }_{0,1}}\left( 1-{{\varepsilon }_{i,0}}-{{\varepsilon }_{i,1}} \right) \right] \right\}}.
\end{aligned}$}
\end{eqnarray}
For the denominator given in (\ref{math49}), we can have that
\begin{eqnarray}\label{math50}
  \left\{
    \begin{aligned}
&\varepsilon_{i,0} < \varepsilon_{i,0} + \pi_{0,1}(1-{{\varepsilon }_{i,0}}-{{\varepsilon }_{i,1}}) < \dfrac{1}{2}, \\
&\dfrac{1}{2} < 1 - \bigl[\varepsilon_{i,1} + \pi_{0,1}(1-{{\varepsilon }_{i,0}}-{{\varepsilon }_{i,1}})\bigr] < 1 - \varepsilon_{i,1}.
\end{aligned}
\right.
\end{eqnarray}

Then we can immediately get that $\frac{\partial \Lambda \left( {{y}_{i}} \right)}{\partial {{\pi }_{0,1}}}<0$.
With the above result, we can easily obtain that since the decision fusion statistic $\Lambda(y_i)$ is positive and monotonically decreases as ${{\pi }_{0,1}}$ increase, its magnitude \(\left| \Lambda(y_i) \right|\) is minimized when ${{\pi }_{0,1}}$ reach maximum, i.e., $\pi_{0,1} = \pi_{1,0} = \alpha$ (corresponding to $P_{0,1} = P_{1,0} = 1$). 
This magnitude minimization directly corresponds to the lowest decision reliability and thus the optimal attack.
Similarly, the derivative of (\ref{math48}) with respect to ${{\pi }_{0,1}}$ yields
\begin{eqnarray}\label{math51}
  \resizebox{0.9\hsize}{!}{$\begin{aligned}
\frac{\partial \Lambda \left( {{y}_{i}} \right)}{\partial {{\pi }_{0,1}}}=\frac{\left( 1-{{\varepsilon }_{i,0}}-{{\varepsilon }_{i,1}} \right)\left( 1-{{\varepsilon }_{i,0}}+{{\varepsilon }_{i,1}} \right)}{\left[ {{\varepsilon }_{i,1}}+{{\pi }_{0,1}}\left( 1-{{\varepsilon }_{i,0}}-{{\varepsilon }_{i,1}} \right) \right]\left\{ 1-\left[ {{\varepsilon }_{i,0}}+{{\pi }_{0,1}}\left( 1-{{\varepsilon }_{i,0}}-{{\varepsilon }_{i,1}} \right) \right] \right\}}.
\end{aligned}$}
\end{eqnarray}

Then we can get that $\frac{\partial \Lambda \left( {{y}_{i}} \right)}{\partial {{\pi }_{0,1}}}>0$. Considering that the decision fusion statistic $\Lambda(y_i)$ is a negative, monotonically increase as ${{\pi }_{0,1}}$ increase, and its magnitude \(\left| \Lambda(y_i) \right|\) is minimized when when ${{\pi }_{0,1}}$ reach maximum. This configuration, i.e., $\pi_{0,1} = \pi_{1,0} = \alpha$, achieves the optimal attack by minimizing the decision reliability.

\subsection{Optimal Small-Scale Byzantine Attack when the SNR for the Multihop Relay Network is High}
For any given ${{P}_{{{D}_{i}}}}$, ${{P}_{{{F}_{i}}}}$, ${{\varepsilon }_{i,0}}$ and ${{\varepsilon }_{i,1}}$, provided that ${{\pi }_{0,1}} = {{\pi }_{1,0}}$, (\ref{math20}) can be rewritten as
\begin{eqnarray}\label{math52}
  \begin{aligned}
\Lambda \left( {{y}_{i}} \right)=\ln \frac{{{P}_{{{D}_{i}}}}-\left( 2{{P}_{{{D}_{i}}}}-1 \right){{\pi }_{0,1}}}{{{P}_{{{F}_{i}}}}+\left( 1-2{{P}_{{{F}_{i}}}} \right){{\pi }_{0,1}}},
\end{aligned}
\end{eqnarray}
and (\ref{math21}) can be changed to
\begin{eqnarray}\label{math53}
  \begin{aligned}
\Lambda \left( {{y}_{i}} \right)=\ln \frac{1-\left[ {{P}_{{{D}_{i}}}}-\left( 2{{P}_{{{D}_{i}}}}-1 \right){{\pi }_{0,1}} \right]}{1-\left[ {{P}_{{{F}_{i}}}}-\left( 2{{P}_{{{F}_{i}}}}-1 \right){{\pi }_{0,1}} \right]}.
\end{aligned}
\end{eqnarray}

The derivative of (\ref{math52}) with respect to ${{\pi }_{0,1}}$ yields
\begin{eqnarray}\label{math54}
  \resizebox{0.9\hsize}{!}{$\begin{aligned}
\frac{\partial \Lambda \left( {{y}_{i}} \right)}{\partial {{\pi }_{0,1}}}=-\left( \frac{2{{P}_{{{D}_{i}}}}-1}{{{P}_{{{D}_{i}}}}\left( 1-2{{\pi }_{0,1}} \right)+{{\pi }_{0,1}}}+\frac{1-2{{P}_{{{F}_{i}}}}}{{{P}_{{{F}_{i}}}}\left( 1-2{{\pi }_{0,1}} \right)+{{\pi }_{0,1}}} \right).
\end{aligned}$}
\end{eqnarray}
Considering that
\begin{eqnarray}\label{math55}
  \resizebox{0.45\hsize}{!}{$\begin{aligned}
\left\{ \begin{aligned}
 & 2{{P}_{{{D}_{i}}}}-1>0, \\ 
 & 1-2{{P}_{{{F}_{i}}}}>0, \\ 
 & {{P}_{{{D}_{i}}}}\left( 1-2{{\pi }_{0,1}} \right)+{{\pi }_{0,1}}>0, \\ 
 & {{P}_{{{F}_{i}}}}\left( 1-2{{\pi }_{0,1}} \right)+{{\pi }_{0,1}}>0,\\ 
\end{aligned} \right.
\end{aligned}$}
\end{eqnarray}
then we can have that $\frac{\partial \Lambda \left( {{y}_{i}} \right)}{\partial {{\pi }_{0,1}}}<0$.
For small-scale attacks, analysis shows that the fusion statistic $\Lambda({y_i})$ is a positive and monotonically decreasing function with respect to ${{\pi }_{0,1}}$. 
Therefore, its magnitude \(\left| \Lambda(y_i) \right|\) is minimized when ${{\pi }_{0,1}}$ reach maximum, i.e., $\pi_{0,1} = \pi_{1,0} = \alpha$.
Similarly, taking the derivative of (\ref{math53}) with respect to \({{\pi }_{0,1}}\) yields
\begin{eqnarray}\label{math56}
  \resizebox{0.9\hsize}{!}{$\begin{aligned}
\frac{\partial \Lambda \left( {{y}_{i}} \right)}{\partial {{\pi }_{0,1}}}=\frac{2{{P}_{{{D}_{i}}}}-1}{{{P}_{{{D}_{i}}}}\left( 1-2{{\pi }_{0,1}} \right)+{{\pi }_{0,1}}}+\frac{1-2{{P}_{{{F}_{i}}}}}{{{P}_{{{F}_{i}}}}\left( 1-2{{\pi }_{0,1}} \right)+{{\pi }_{0,1}}}.
\end{aligned}$}
\end{eqnarray}

As indicated in (\ref{math56}), we can still easily derive that when \({\pi_{0,1}} = {\pi_{1,0}} = \alpha\), the magnitude \(\left| \Lambda(y_i) \right|\) is minimized, leading to the lowest reliability of \(\Lambda(y_i)\).

\subsection{Optimal Small-Scale Byzantine Attack when the SNR for the Multihop Relay Network is Low}
In this context, if substituting ${{\pi }_{0,1}}={{\pi }_{1,0}}$ into (\ref{math25}) and (\ref{math26}), we can respectively have
\begin{eqnarray}\label{math57}
  \begin{aligned}
\Lambda \left( {{y}_{i}} \right)=\left( 1-2{{\pi }_{0,1}} \right)\left( {{P}_{{{D}_{i}}}}-{{P}_{{{F}_{i}}}} \right)\underset{1\le j\le J}{\mathop{\min }}\,\left( {{\delta }_{i,j}} \right),
\end{aligned}
\end{eqnarray}
and
\begin{eqnarray}\label{math58}
  \begin{aligned}
\Lambda \left( {{y}_{i}} \right)\approx -\left( 1-2{{\pi }_{0,1}} \right)\left( {{P}_{{{D}_{i}}}}-{{P}_{{{F}_{i}}}} \right)\underset{1\le j\le J}{\mathop{\min }}\,\left( {{\delta }_{i,j}} \right).
\end{aligned}
\end{eqnarray}
Further, if both \({\pi}_{0,1}\) and \({\pi}_{1,0}\) reach their maximum \(\alpha\), it is straightforward to observe that \((1 - 2{\pi}_{0,1})\) attains its minimum. Consequently, (\ref{math57}) and (\ref{math58}) respectively reach their minimum and maximum. This implies that the magnitude of the fusion statistic \(\Lambda(y_i)\) is minimized, thereby leading to the lowest possible reliability of \(\Lambda(y_i)\).

\subsection{Optimal Large-Scale Byzantine Attack with Ideal Local Channel}
For any given $P_{D_i}$, $P_{F_i}$, ${{\varepsilon}_{i,0}}$ and ${{\varepsilon}_{i,1}}$, if $\pi_{0,1}$ and $\pi_{1,0}$ satisfy $\pi_{0,1} + \pi_{1,0} = 1$, then (\ref{math18}) can be revised as
\begin{align}\label{math59} 
\Lambda \left( {{y}_{i}} \right)=\ln \frac{1-{{\varepsilon}_{i,1}}-{{\pi }_{0,1}}\left( 1-{{\varepsilon}_{i,0}}-{{\varepsilon}_{i,1}} \right)}{{{\varepsilon }_{i,0}}+\left( 1-{{\pi }_{0,1}} \right)\left( 1-{{\varepsilon}_{i,0}}-{{\varepsilon}_{i,1}} \right)}=0,
\end{align}
and (\ref{math19}) can be revised as
\begin{eqnarray}\label{math60}
  \begin{aligned}
\Lambda \left( {{y}_{i}} \right)=\ln \frac{{{\varepsilon }_{i,0}}+{{\pi }_{0,1}}\left( 1-{{\varepsilon}_{i,0}}-{{\varepsilon}_{i,1}} \right)}{1-{{\varepsilon }_{i,1}}-\left( 1-{{\pi }_{0,1}} \right)\left( 1-{{\varepsilon}_{i,0}}-{{\varepsilon}_{i,1}} \right)}=0.
\end{aligned}
\end{eqnarray}

It is evident that for any given ${{P}_{{D}_{i}}}$, ${{P}_{{F}_{i}}}$, ${\varepsilon}_{i,0}$, and ${\varepsilon}_{i,1}$, the magnitude of the fusion statistic $\Lambda(y_i)$ is always 0, indicating its reliability is minimized.

\subsection{Optimal Large-Scale Byzantine Attack when the SNR for the Multihop Relay Network is High}
If we bring ${{\pi }_{0,1}}+{{\pi }_{1,0}}=1$ into the suboptimal fusion rule (\ref{math20}) and (\ref{math21}), we can respectively have
\begin{eqnarray}\label{math61}
  \begin{aligned}
\Lambda \left( {{y}_{i}} \right)=\ln \frac{{{P}_{{{D}_{i}}}}\left( 1-{{\pi }_{0,1}} \right)+\left( 1-{{P}_{{{D}_{i}}}} \right){{\pi }_{1,0}}}{{{P}_{{{F}_{i}}}}\left( 1-{{\pi }_{0,1}} \right)+\left( 1-{{P}_{{{F}_{i}}}} \right){{\pi }_{1,0}}}=0,
\end{aligned}
\end{eqnarray}
and
\begin{eqnarray}\label{math62}
  \begin{aligned}
\Lambda \left( {{y}_{i}} \right)=\ln \frac{{{P}_{{{D}_{i}}}}{{\pi }_{0,1}}+\left( 1-{{P}_{{{D}_{i}}}} \right)\left( 1-{{\pi }_{1,0}} \right)}{{{P}_{{{F}_{i}}}}{{\pi }_{0,1}}+\left( 1-{{P}_{{{F}_{i}}}} \right)\left( 1-{{\pi }_{1,0}} \right)}=0.
\end{aligned}
\end{eqnarray}
It is evident that for any given ${{P}_{{D}_{i}}}$, ${{P}_{{F}_{i}}}$ ${\varepsilon }_{i,0}$ and ${\varepsilon }_{i,1}$ the magnitude of the fusion statistic $\Lambda \left( {{y}_{i}} \right)$ is always 0, indicating that the reliability of the fusion statistic is minimized.

\subsection{Optimal Large-Scale Byzantine Attack when the SNR for the Multihop Relay Network is Low}
If we bring ${{\pi }_{0,1}}+{{\pi }_{1,0}}=1$ into (\ref{math25}) and (\ref{math26}), we can respectively have
\begin{eqnarray}\label{math63}
  \begin{aligned}
\Lambda \left( {{y}_{i}} \right)=\left( 1-{{\pi }_{0,1}}-{{\pi }_{1,0}} \right)\left( {{P}_{{{D}_{i}}}}-{{P}_{{{F}_{i}}}} \right)\underset{1\le j\le J}{\mathop{\min }}\,\left( {{\delta }_{i,j}} \right)=0,
\end{aligned}
\end{eqnarray}
and
\begin{eqnarray}\label{math64}
  \resizebox{0.9\hsize}{!}{$\begin{aligned}
\Lambda \left( {{y}_{i}} \right)\approx -\left( 1-{{\pi }_{0,1}}-{{\pi }_{1,0}} \right)\left( {{P}_{{{D}_{i}}}}-{{P}_{{{F}_{i}}}} \right)\underset{1\le j\le J}{\mathop{\min }}\,\left( {{\delta }_{i,j}} \right)=0.
\end{aligned}$}
\end{eqnarray}
Similarly, at this point, the decision fusion statistic $\Lambda \left( {{y}_{i}} \right)$ is 0 for any given ${{P}_{{D}_{i}}}$, ${{P}_{{F}_{i}}}$, ${{\varepsilon }_{i,0}}$ and ${{\varepsilon }_{i,1}}$.

The above interpretation cognizant of the fact that the principle of the Byzantine attack is to minimize the magnitude of the decision fusion statistic $\Lambda(y_i)$, and this magnitude minimization could significantly reduce the reliability of the information gathered from local SUs by the FC. This is not a coincidence: the attackers is to make the data that the FC receives from the SUs such that no information is conveyed. 
Thus, the FC becomes blind if the conditional probabilities given in (\ref{math13})  are identical. 
In such a scenario, the best that the FC can do is to make decisions solely based on the priors, resulting in the most degraded performance at the FC.
Based on the above discussion, we derive optimal Byzantine attack strategies for both large-scale and small-scale scenarios, as summarized in Algorithm \ref{alg:byzantine_attack}.
\begin{algorithm}
  \caption{Optimal Byzantine Attack Algorithm}
  \label{alg:byzantine_attack}
  \begin{algorithmic}[1]
    \STATE \textbf{Input:} $I$ (number of SUs); $x$ (PU status after SU decision, $x_i\in\{0,1\}$); 
    \STATE \hspace{1.1em} $\alpha$ (attack intensity); $P_{0,1}$; $P_{1,0}$.
    \STATE \textbf{Output:} $u$ (attack result)
    \IF{$\alpha \le 0.5$}
      \STATE \textit{(AF attack)}
      \FOR{$i=1$ \TO $I$}
        \IF{$i \in \text{Byzantine\_nodes}$}
          \STATE $u_i \leftarrow x_i \oplus 1$
        \ELSE
          \STATE $u_i \leftarrow x_i$
        \ENDIF
      \ENDFOR
    \ELSE
      \STATE \textit{(satisfy $\alpha(P_{0,1}+P_{1,0})=1$)}
      \FOR{$i=1$ \TO $I$}
        \IF{$i \in \text{Byzantine\_nodes}$}
          \STATE $u_i \leftarrow \mathrm{BC}(P_{0,1},P_{1,0},x_i)$
        \ELSE
          \STATE $u_i \leftarrow x_i$
        \ENDIF
      \ENDFOR
    \ENDIF
    \STATE \textbf{return} $u$
  \end{algorithmic}
\end{algorithm}

\section{Discussion}\label{section7}
As shown in the above analysis of the optimal Byzantine attack under both optimal and suboptimal decision fusion rules, we find that the attacker can achieve the optimal attack effect through a specific attack strategy regardless of whether the decision fusion rule is known or not. The key observations are summarized as follows:

\begin{enumerate}
    \item For small-scale Byzantine attacks, the optimal attack strategy is the AF attack, i.e., flipping the local spectrum sensing report so as to minimize the magnitude of the decision fusion statistic \(\Lambda(y_i)\) from the attacker perspective. This magnitude reduction degrades the global decision reliability of the FC.
    \item For large-scale Byzantine attacks, the optimal attack strategy is to flip local spectrum sensing report extensively such that the decision fusion statistic \(\Lambda(y_i)\) is driven to 0. In this context, the FC becomes incapable of recovering any meaningful information from the corrupted received spectrum sensing reports, resulting in a complete failure of its decision-making. 
    \item The attacker can achieve optimal impact even when no tangible ICSI is available, simply by selecting an appropriate strategy based on the fraction of the Byzantine nodes. Thus, the
reliance on the global ICSI and decision fusion rules in obtaining the Byzantine attacks can be relaxed, and the Byzantine attacks can be successfully implemented over time-varying mobile channels without any a prior knowledge of decision fusion rules or ICSI.
\end{enumerate}

\section{Attack performance evaluation}\label{section8}
\noindent According to our previous analysis of optimal Byzantine attack strategies and their impact on decision fusion, this section conducts a systematic evaluation of typical attack strategies.
Based on their characteristics, these attacks can be categorized into the following four types \cite{15}:

\begin{enumerate}
  \item AN attack: data 0 remains unchanged, data 1 is flipped to 0;

  \item AY attack: data 1 remains unchanged, data 0 is flipped to 1;

  \item AF attack: data 0 is flipped to 1, data 1 is flipped to 0;

  \item RD attack: data 0 is randomly flipped to 1, data 1 is randomly flipped to 0.
\end{enumerate}

\subsection{Performance Evaluation when ${{y}_{i}}=1$}
\noindent In order to evaluate the overall attack performance we first evaluate the performance of individual branches, if the attack performance of each branch is consistent, it can also indicate the strength of the attack performance of different attack methods on the overall system.

When ${{y}_{i}}=1$, (\ref{math31}) is reduced to
\begin{eqnarray}\label{math72}
  \small
  \begin{aligned}
    \Lambda \left( {{y}_{i}} \right)=\ln \frac{A}{B},
\end{aligned}
\end{eqnarray}
where $A$ and $B$ are
\begin{equation}\label{math73}
  \small
  \left\{
    \resizebox{0.8\hsize}{!}{$\begin{aligned}
    A &= {{P}_{{{D}_{i}}}}\left( 1-{{\varepsilon }_{i,1}} \right)+\left( 1-{{P}_{{{D}_{i}}}} \right){{\varepsilon }_{i,0}} \\
      &\quad -\left[ {{P}_{{{D}_{i}}}}{{\pi }_{0,1}}-\left( 1-{{P}_{{{D}_{i}}}} \right){{\pi }_{1,0}} \right]\left( 1-{{\varepsilon }_{i,0}}-{{\varepsilon }_{i,1}} \right) \\ 
    B &= {{P}_{{{F}_{i}}}}\left( 1-{{\varepsilon }_{i,1}} \right)+\left( 1-{{P}_{{{F}_{i}}}} \right){{\varepsilon }_{i,0}} \\
      &\quad -\left[ {{P}_{{{F}_{i}}}}{{\pi }_{0,1}}-\left( 1-{{P}_{{{F}_{i}}}} \right){{\pi }_{1,0}} \right]\left( 1-{{\varepsilon }_{i,0}}-{{\varepsilon }_{i,1}} \right). 
  \end{aligned}$}
  \right.
\end{equation}

The four attack modes can be represented as follows.

AN attack:
\begin{eqnarray}\label{math74}
  \small
  \begin{aligned}
\left\{ \begin{aligned}
  & {{A}_{AN}}={{\varepsilon }_{i,0}}+{{P}_{{{D}_{i}}}}\left( 1-{{\varepsilon }_{i,0}}-{{\varepsilon }_{i,1}} \right)\left( 1-\alpha  \right) \\ 
 & {{B}_{AN}}={{\varepsilon }_{i,0}}+{{P}_{{{F}_{i}}}}\left( 1-{{\varepsilon }_{i,0}}-{{\varepsilon }_{i,1}} \right)\left( 1-\alpha  \right). \\ 
\end{aligned} \right.
\end{aligned}
\end{eqnarray}

AY attack:
\begin{eqnarray}\label{math75}
  \small
  \resizebox{0.9\hsize}{!}{$\begin{aligned}
\left\{ \begin{aligned}
  & {{A}_{AY}}={{\varepsilon }_{i,0}}+\left[ \alpha +\left( 1-\alpha  \right){{P}_{{{D}_{i}}}} \right]\left( 1-{{\varepsilon }_{i,0}}-{{\varepsilon }_{i,1}} \right) \\ 
 & {{B}_{AY}}={{\varepsilon }_{i,0}}+\left[ \alpha +\left( 1-\alpha  \right){{P}_{{{F}_{i}}}} \right]\left( 1-{{\varepsilon }_{i,0}}-{{\varepsilon }_{i,1}} \right). \\ 
\end{aligned} \right.
\end{aligned}$}
\end{eqnarray}

AF Attack:
\begin{eqnarray}\label{math76}
  \small
  \resizebox{0.9\hsize}{!}{$\begin{aligned}
\left\{ \begin{aligned}
  & {{A}_{AF}}={{\varepsilon }_{i,0}}+\left\{ \alpha -\left( 2\alpha -1 \right){{P}_{{{D}_{i}}}} \right\}\left( 1-{{\varepsilon }_{i,0}}-{{\varepsilon }_{i,1}} \right) \\ 
 & {{B}_{AF}}={{\varepsilon }_{i,0}}+\left\{ \alpha -\left( 2\alpha -1 \right){{P}_{{{F}_{i}}}} \right\}\left( 1-{{\varepsilon }_{i,0}}-{{\varepsilon }_{i,1}} \right). \\ 
\end{aligned} \right.
\end{aligned}$}
\end{eqnarray}

RF Attack:
\begin{eqnarray}\label{math77}
  \small
  \resizebox{0.9\hsize}{!}{$\begin{aligned}
\left\{ \begin{aligned}
  &{{A}_{RD}}={{\varepsilon }_{i,0}}+\left[ {{P}_{{{D}_{i}}}}+{{P}_{1,0}}-{{P}_{{{D}_{i}}}}\alpha ({{P}_{0,1}}+{{P}_{1,0}}) \right]\left( 1-{{\varepsilon }_{i,0}}-{{\varepsilon }_{i,1}} \right)\\ 
 &{{B}_{RD}}={{\varepsilon }_{i,0}}+\left[ {{P}_{{{D}_{i}}}}+{{P}_{1,0}}-{{P}_{{{F}_{i}}}}\alpha ({{P}_{0,1}}+{{P}_{1,0}}) \right]\left( 1-{{\varepsilon }_{i,0}}-{{\varepsilon }_{i,1}} \right). \\ 
\end{aligned} \right.
\end{aligned}$}
\end{eqnarray}

\subsubsection{AN Attacks and AY Attacks}
For the AN attack, the numerator is approximately proportional to:
$P_{D_i} (1 - \alpha)$.
The denominator is approximately proportional to:
$P_{F_i} (1 - \alpha)$.
Therefore, the overall decision statistic is proportional to:
$\left( P_{D_i} - P_{F_i} \right) (1 - \alpha)$.

Similarly, for the AY attack, the numerator is approximately proportional to:
$\alpha + (1 - \alpha) P_{D_i}$.
The denominator is approximately proportional to:
$\alpha + (1 - \alpha) P_{F_i}$.
Therefore, the overall decision statistic is proportional to:
$\left( P_{D_i} - P_{F_i} \right) (1 - \alpha)$.
Thus, both the AN and AY attacks have the same overall decision statistic, proportional to: $\left( P_{D_i} - P_{F_i} \right) (1 - \alpha)$.
Therefore, whether small or large-scale, the attack performance of AN and AY attacks is considered identical.

\subsubsection{AF Attacks and AN Attacks}
According to (\ref{math76}), for the AF attack, the numerator is approximately proportional to: $\left[ \alpha - (2\alpha - 1) P_{D_i} \right]$. The denominator is approximately proportional to: $\left[ \alpha - (2\alpha - 1) P_{F_i} \right]$. Therefore, the overall decision statistic is proportional to: $\left( P_{D_i} - P_{F_i} \right) (2\alpha - 1)$.

Since it is known that \( P_{D_i} - P_{F_i} > 0 \), comparing \( (1 - \alpha) \) and \( |2\alpha - 1| \) can reveal which attack results in a smaller magnitude for \( \Lambda(y_i) \).

\begin{itemize}
  \item \textbf{Small-scale attack}: When \( \alpha < 0.5 \), it can be concluded that under small-scale attacks, the AF attack outperforms the AN attack.
  \item \textbf{Large-scale attack}: When \( \frac{1}{2} \le \alpha \le \frac{2}{3} \), the AF attack performs better. When \( \frac{2}{3} \le \alpha \le 1 \), the AN or AY attacks perform better.
\end{itemize}

\subsubsection{ RF Attack and AN Attack}
For the RF attack, the numerator is approximately proportional to:

$\left[ P_{D_i} + \alpha P_{1,0} - P_{D_i} \alpha (P_{0,1} + P_{1,0}) \right]$.

The denominator is approximately proportional to:

$\left[ P_{F_i} + \alpha P_{1,0} - P_{F_i} \alpha (P_{0,1} + P_{1,0}) \right]$.

Therefore, the overall decision statistic is proportional to:

$\left( P_{D_i} - P_{F_i} \right) \left[ \alpha (P_{0,1} + P_{1,0}) - 1 \right]$.

Performance comparison of RD and AN Attacks, we can use \( \left| \alpha (P_{0,1} + P_{1,0}) - 1 \right| \) and \( (1 - \alpha) \) to determine which attack results in a smaller magnitude for \( \Lambda(y_i) \).

\begin{itemize}
  \item \textbf{Small-scale attack}: Since \( \alpha < 0.5 \), comparing \( 1 - \alpha (P_{0,1} + P_{1,0}) \) and \( (1 - \alpha) \), we can conclude that: When \( 0 \le P_{0,1} + P_{1,0} \le 1 \), the AN attack performs better. When \( 1 \le P_{0,1} + P_{1,0} \le 2 \), the RD attack performs better.  
  \item \textbf{Large-scale attack}: We have \( \left| \alpha (P_{0,1} + P_{1,0}) - 1 \right| = \alpha (P_{0,1} + P_{1,0}) - 1 \). When \( 0 \le P_{0,1} + P_{1,0} \le 1 \), it is concluded that \( (1 - \alpha) \ge \left| \alpha (P_{0,1} + P_{1,0}) - 1 \right| \), meaning that the performance of AN or AY attacks is always better than that of the AF attack. When \( 1 \le P_{0,1} + P_{1,0} \le 2 \), setting \( (1 - \alpha) = \alpha (P_{0,1} + P_{1,0}) - 1 \), we get:
$P_{0,1} + P_{1,0} = \frac{2}{\alpha} - 1$. When \( 1 \le P_{0,1} + P_{1,0} \le \frac{2}{\alpha} - 1 \), the RD attack performs better. When \( \frac{2}{\alpha} - 1 \le P_{0,1} + P_{1,0} \le 2 \), the AN attack performs better.
\end{itemize}

\subsubsection{ RD Attack and AF Attack}
To compare the performance of the RD attack with the AF attack, we can use \( \left| \alpha (P_{0,1} + P_{1,0}) - 1 \right| \) and \( (2\alpha - 1) \) to determine which attack has better performance.

\begin{itemize}
  \item \textbf{Small-scale attack}: When \( 0 \le P_{0,1} + P_{1,0} \le 1 \), since AF attacks require $P_{0,1}=P_{1,0} =1$, the AF attack does not exist at this point; when \( 1 \le P_{0,1} + P_{1,0} \le 2 \), the AF attack performs better.
  \item \textbf{Large-scale attack}: When \( 1 \le P_{0,1} + P_{1,0} \le \frac{2}{\alpha} - 2 \), the AF attack performs better; when \( \frac{2}{\alpha} - 2 \le P_{0,1} + P_{1,0} \le 2 \), the RD attack performs better.
\end{itemize}

\subsection{Performance Evaluation when ${{y}_{i}}=0$}

\noindent When ${{y}_{i}}=0$, (\ref{math32}) can be reduced to
\begin{small}
\begin{eqnarray}\label{math78}
  \begin{aligned}
\Lambda \left( {{y}_{i}} \right)=\ln \frac{C}{D},
\end{aligned}
\end{eqnarray}
\end{small}
where $C$ and $D$ are
\begin{small}
\begin{eqnarray}\label{math79}
  \resizebox{0.8\hsize}{!}{$\begin{aligned}
\left\{ \begin{aligned}
  C&={{P}_{{{D}_{i}}}}{{\varepsilon }_{i,1}}+\left( 1-{{P}_{{{D}_{i}}}} \right)\left( 1-{{\varepsilon }_{i,0}} \right)\\
  &\quad+\left[ {{P}_{{{D}_{i}}}}{{\pi }_{0,1}}-\left( 1-{{P}_{{{D}_{i}}}} \right){{\pi }_{1,0}} \right]\left( 1-{{\varepsilon }_{i,0}}-{{\varepsilon }_{i,1}} \right) \\ 
  D&={{P}_{{{F}_{i}}}}{{\varepsilon }_{i,1}}+\left( 1-{{P}_{{{F}_{i}}}} \right)\left( 1-{{\varepsilon }_{i,0}} \right)\\
  &\quad+\left[ {{P}_{{{F}_{i}}}}{{\pi }_{0,1}}-\left( 1-{{P}_{{{F}_{i}}}} \right){{\pi }_{1,0}} \right]\left( 1-{{\varepsilon }_{i,0}}-{{\varepsilon }_{i,1}} \right). \\ 
\end{aligned} \right.
\end{aligned}$}
\end{eqnarray}
\end{small}

Substituting the four attack patterns at this point yields:

AN attack:
\begin{small}
\begin{eqnarray}\label{math80}
  \resizebox{0.9\hsize}{!}{$\begin{aligned}
\left\{ \begin{aligned}
 & {{C}_{AN}}=\left( 1-{{\varepsilon }_{i,0}} \right)-{{P}_{{{D}_{i}}}}\left( 1-{{\varepsilon }_{i,0}}-{{\varepsilon }_{i,1}} \right)\left( 1-\alpha  \right) \\ 
 &{{D}_{AN}}=\left( 1-{{\varepsilon }_{i,0}} \right)-{{P}_{{{F}_{i}}}}\left( 1-{{\varepsilon }_{i,0}}-{{\varepsilon }_{i,1}} \right)\left( 1-\alpha  \right). \\ 
\end{aligned} \right.
\end{aligned}$}
\end{eqnarray}
\end{small}

AY attack:
\begin{small}
\begin{eqnarray}\label{math81}
  \resizebox{0.9\hsize}{!}{$\begin{aligned}
\left\{ \begin{aligned}
  &{{C}_{AY}}=\left( 1-{{\varepsilon }_{i,0}} \right)-\left[ \alpha +\left( 1-\alpha  \right){{P}_{{{D}_{i}}}} \right]\left( 1-{{\varepsilon }_{i,0}}-{{\varepsilon }_{i,1}} \right) \\ 
 &{{D}_{AY}}=\left( 1-{{\varepsilon }_{i,0}} \right)-\left[ \alpha +\left( 1-\alpha  \right){{P}_{F}} \right]\left( 1-{{\varepsilon }_{i,0}}-{{\varepsilon }_{i,1}} \right). \\ 
\end{aligned} \right.
\end{aligned}$}
\end{eqnarray}
\end{small}

AF attack:
\begin{small}
\begin{eqnarray}\label{math82}
  \resizebox{0.9\hsize}{!}{$\begin{aligned}
    \left\{ \begin{aligned}
&{{C}_{_{AF}}}=\left( 1-{{\varepsilon }_{i,0}} \right)-\left\{ \alpha -\left( 2\alpha -1 \right){{P}_{{{D}_{i}}}} \right\}\left( 1-{{\varepsilon }_{i,0}}-{{\varepsilon }_{i,1}} \right) \\ 
 &{{D}_{_{AF}}}=\left( 1-{{\varepsilon }_{i,0}} \right)-\left\{ \alpha -\left( 2\alpha -1 \right){{P}_{{{F}_{i}}}} \right\}\left( 1-{{\varepsilon }_{i,0}}-{{\varepsilon }_{i,1}} \right). \\ 
\end{aligned} \right.
\end{aligned}$}
\end{eqnarray}
\end{small}

RD attack:
\begin{small}
\begin{eqnarray}\label{math83}
  \resizebox{0.9\hsize}{!}{$\begin{aligned}
    \left\{ \begin{aligned}
&{{C}_{_{RD}}}=\left( 1-{{\varepsilon }_{i,0}} \right)-\left[ {{P}_{{{D}_{i}}}}+\alpha {{P}_{1,0}}-{{P}_{{{D}_{i}}}}\alpha ({{P}_{0,1}}+{{P}_{1,0}}) \right]\left( 1-{{\varepsilon }_{i,0}}-{{\varepsilon }_{i,1}} \right) \\ 
 &{{D}_{_{RD}}}=\left( 1-{{\varepsilon }_{i,0}} \right)-\left[ {{P}_{{{F}_{i}}}}+\alpha {{P}_{1,0}}-{{P}_{{{F}_{i}}}}\alpha ({{P}_{0,1}}+{{P}_{1,0}}) \right]\left( 1-{{\varepsilon }_{i,0}}-{{\varepsilon }_{i,1}} \right). \\ 
\end{aligned} \right.
\end{aligned}$}
\end{eqnarray}
\end{small}

Similar to the analysis when ${{y}_{i}}=1$, the strength of the attack performance of different attack methods can also be obtained for different scales of attack intensity when ${{y}_{i}}=0$. However, the conclusions drawn are consistent with those for ${{y}_{i}}=1$, and the specific analyses are shown in the \text{Appendix A}.

\section{NUMERICAL RESULTS AND DISCUSSIONS}\label{section9}
\begin{table}[!t]
  \renewcommand{\arraystretch}{0.8} 
  \small
  \begin{center}
  \caption{Parameters Used in Simulations }\label{table4}
    \scalebox{1.1}{
  \begin{tabular}{cc}
  \hline
  Parameter                                                                                                                                 &  Detailed description \\ \hline
 \begin{tabular}[c]{@{}c@{}}{Antenna Number for PU}\end{tabular}                                                                            &  1 \\
  Number of SUs $I$                                                                                                                         &  8, 10, 12\\
 \begin{tabular}[c]{@{}c@{}}{Antenna Number}\\ {for each SU $M$}\end{tabular}                                                               &  4, 6, 8 \\
 \begin{tabular}[c]{@{}c@{}}{Number of RIS}\\ {reflective elements  $N$}\end{tabular}                                                       &  5, 7, 9 \\ 
 \begin{tabular}[c]{@{}c@{}}{Channel condition  }\\{$\mathbf{h}_{d_i}$, ${\mathbf{h}_{r}}$, and ${\mathbf{{H}}_{i}}$}\end{tabular}          &\begin{tabular}[c]{@{}c@{}}{Rayleigh fading with}\\ {normalised average power}\end{tabular}\\ 
  Noise condition $ \mathbf{n}_{i}(t)$                                                                                                      & Complex AWGN\\
  Sample number $T$                                                                                                                         & 50\\
  Phase shift of RIS $\bm{\Theta}$                                                                                                          & Randomly generated\\
  Number of hops $J$                                                                                                                        &4, 6, 8\\
  Information sequence length                                                                                                               &504 bits\\
  Number of simulation cycles                      &\begin{tabular}[c]{@{}c@{}}{Get at least 3000}\\ {frame errors}\end{tabular}  \\  \hline
  \end{tabular}}
  \end{center}
\end{table}

This study assesses the influence of Byzantine attacks on system performance, quantified by Bit Error Rate (BER), within time-varying channel environments. Our analysis systematically investigates six key factors, such as the proportion of Byzantine attackers, the number of antennas per SU, and the number of reflecting elements in the RIS, to determine optimal attack strategies for both small-scale and large-scale scenarios. The default simulation parameters are as follows: 10 SUs each equipped with 6 antennas, an RIS with 9 elements, 8 relay nodes, and a data sequence length of 504 bits. A complete list of parameters is provided in Table \ref{table4}.

\begin{figure}[t]
  \centering
  \includegraphics[width=0.65\linewidth]{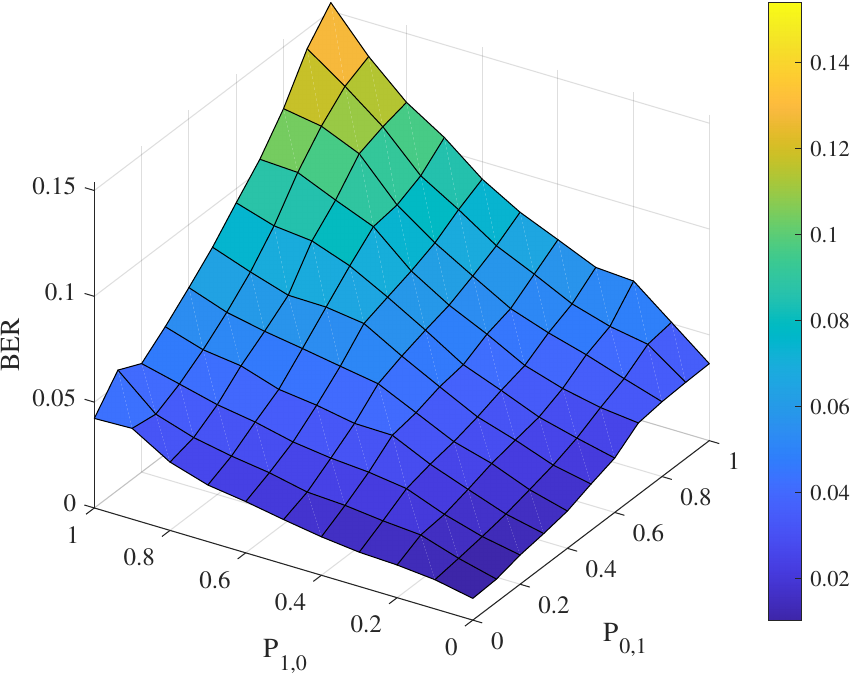}
  \captionsetup{skip=4pt}
  \caption{The BER Performance of Optimal Fusion Rule under Small-Scale Attacks.}
  \label{fig4}
\end{figure} 

\subsection{The Strategy Analysis of Small-Scale Byzantine Attacks under the Optimal Fusion Rule}
\begin{figure}[t]
  \centering
  \includegraphics[width=0.65\linewidth]{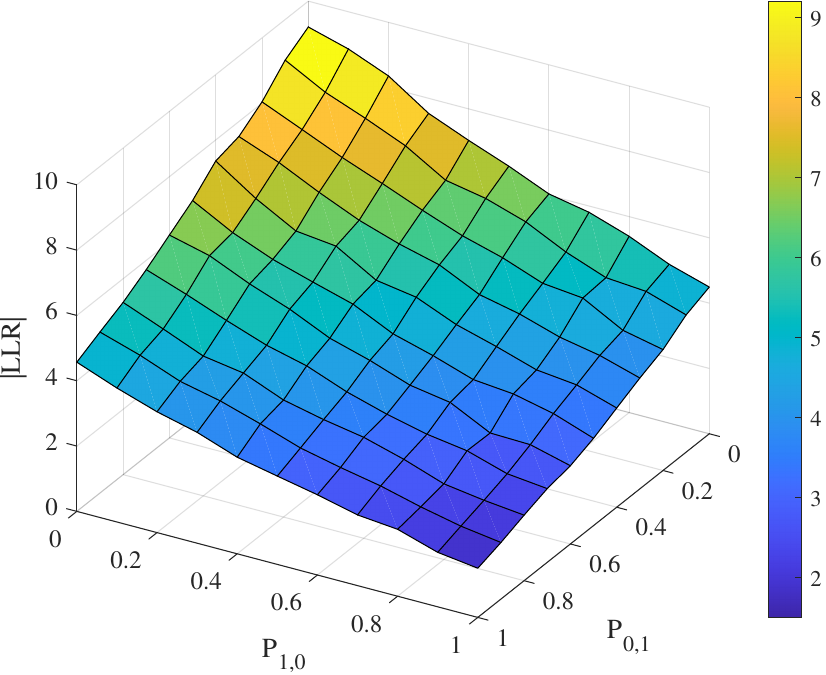}
  \captionsetup{skip=4pt}
  \caption{The Magnitude of LLR for Optimal Fusion Rules under Small-scale Attacks.}
  \label{fig4-1}
\end{figure}

To substantiate the previously derived optimal strategy for small-scale Byzantine attacks under the optimal fusion rule, we conducted targeted simulations using the default parameter configuration for the optimal small-scale scenario. This verification aims to confirm whether the theoretical predictions hold under practical settings and to quantify the performance degradation caused by such attacks. The BER and the Magnitude of the LLR were adopted as key performance metrics, as they directly reflect decision accuracy and detection confidence. Simulation results reveal that, under optimal attack conditions, the BER attains its maximum value, which is shown in Fig.\ref{fig4}, while the Magnitude of LLR reaches its minimum, which is shown in Fig.\ref{fig4-1}, indicating severe impairment of detection reliability.
In contrast, without attacks, the BER is minimized and the Magnitude of LLR is maximized, reflecting stable and reliable system operation. These outcomes not only validate the theoretical findings but also underscore the critical vulnerability of the system to optimally configured small-scale Byzantine attacks.

\subsection{The Robustness Evaluation of Small-Scale Byzantine Attack Strategies under Varying Parameters}
\begin{figure*}[t]
  \centering
  \begin{subfigure}[t]{0.24\textwidth}
    \includegraphics[width=\linewidth]{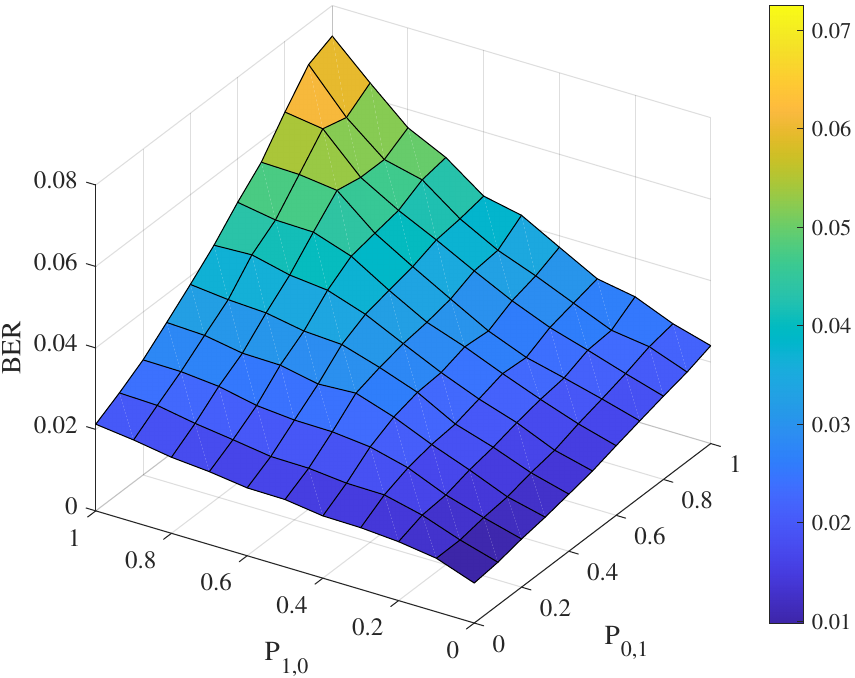}
    \captionsetup{skip=4pt}
    \caption{\footnotesize $\alpha=0.2$} 
    \label{fig:sub51}
\end{subfigure}
  \hfill
\begin{subfigure}[t]{0.24\textwidth}
    \includegraphics[width=\linewidth]{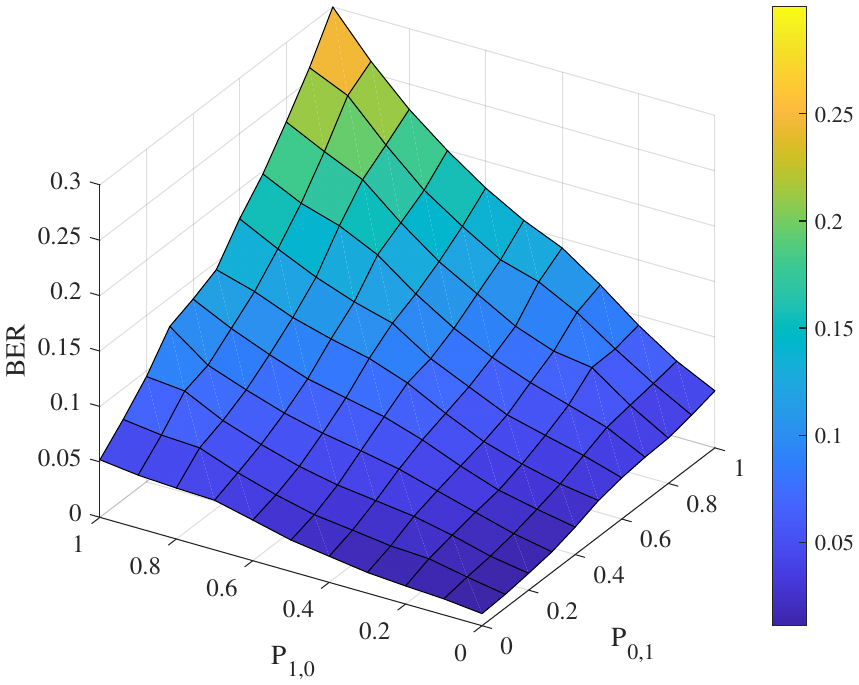}
    \captionsetup{skip=4pt}
    \caption{\footnotesize $\alpha=0.4$}
    \label{fig:sub52}
\end{subfigure}
  \hfill
\begin{subfigure}[t]{0.24\textwidth}
    \includegraphics[width=\linewidth]{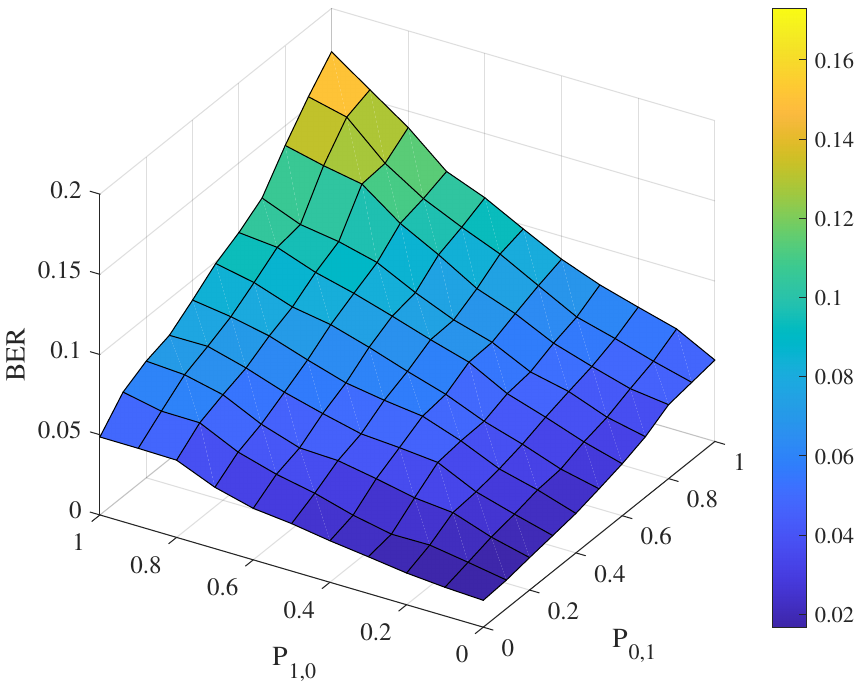}
    \captionsetup{skip=4pt}
    \caption{\footnotesize $M=4$}
    \label{fig:sub53}
\end{subfigure}
\begin{subfigure}[t]{0.24\textwidth}
    \includegraphics[width=\linewidth]{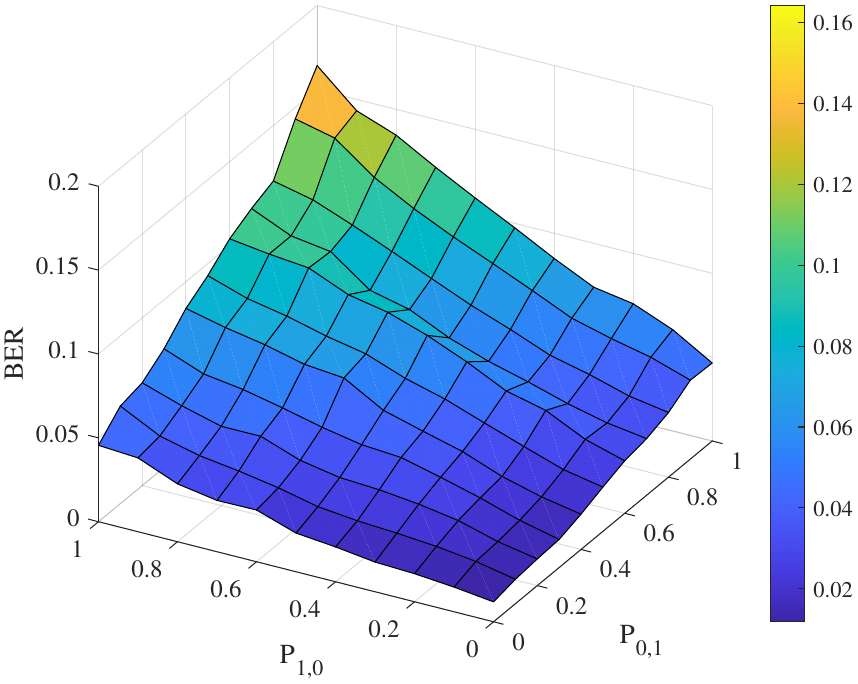}
    \captionsetup{skip=4pt}
    \caption{\footnotesize $M=8$ }
    \label{fig:sub54}
  \end{subfigure}
  \captionsetup{skip=4pt}
  \caption{The BER performance of Optimal Fusion Rule under different Attack Strengths and Numbers of SUs Antennas.}
  \label{fig5}
\end{figure*}

Based on theoretical derivations, this study concludes that in wireless networks, the effectiveness of the optimal small-scale Byzantine attack depends only on the proportion of Byzantine nodes and the signal flipping probability, and is independent of CSI. Accordingly, a multidimensional evaluation framework was established to analyze system performance from three aspects: the number of antennas per SU, the total number of SUs, and the number of RIS reflecting elements. The BER under various parameter settings was used to quantify the effectiveness of different attack strategies.
Building upon the previous analysis, we further simulated the system by varying four key channel-related parameters. As shown in Fig. \ref{fig5}, increasing the number of antennas significantly improves overall performance, with smoother BER curves indicating enhanced anti-interference capability.

\begin{figure*}[t]
  \centering
  \begin{subfigure}[t]{0.24\textwidth}
    \includegraphics[width=\linewidth]{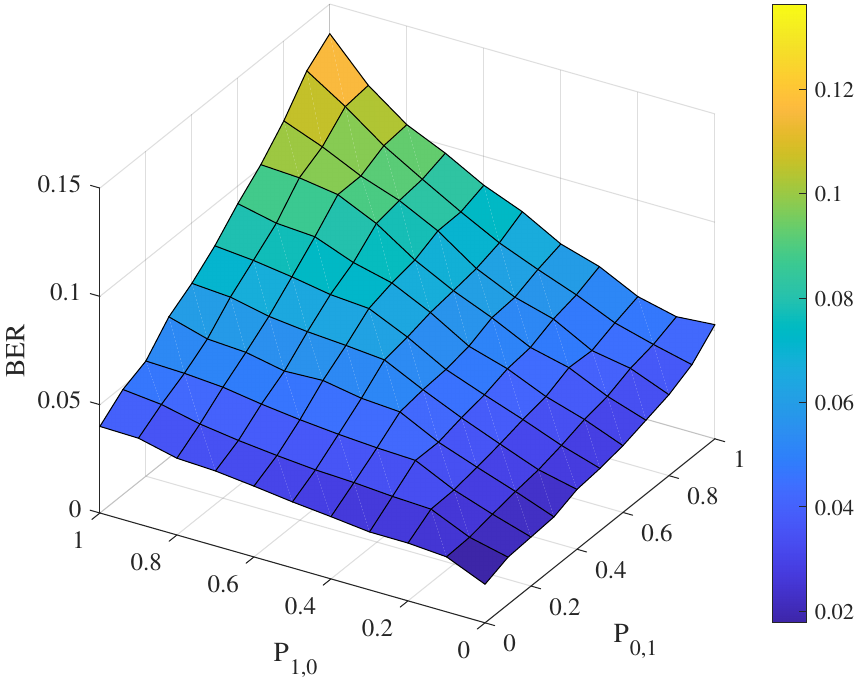}
    \captionsetup{skip=4pt}
    \caption{\footnotesize $I=8$} 
    \label{fig:sub61}
  \end{subfigure}
  \hfill
  \begin{subfigure}[t]{0.24\textwidth}
    \includegraphics[width=\linewidth]{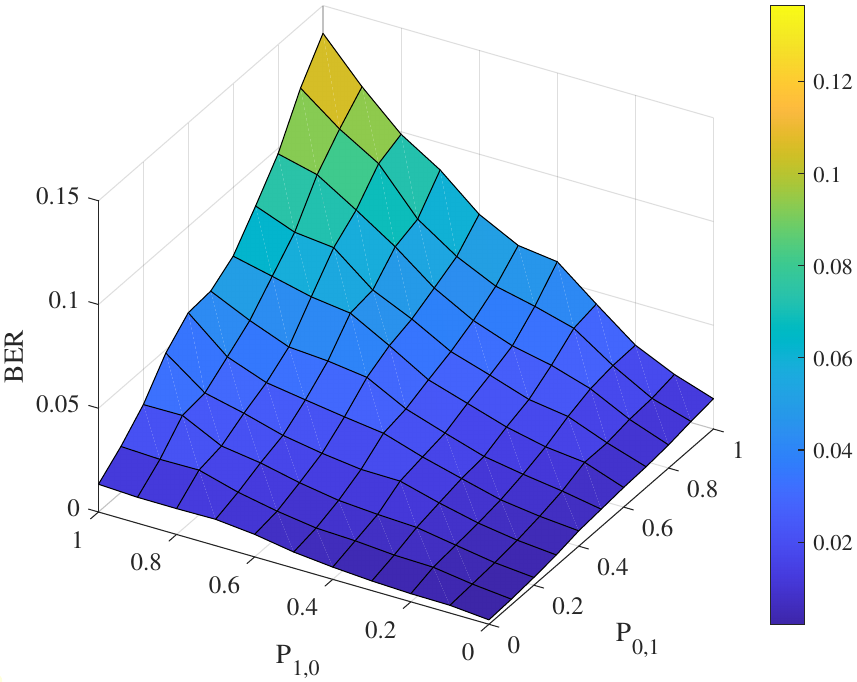}
    \captionsetup{skip=4pt}
    \caption{\footnotesize $I=12$}
    \label{fig:sub62}
  \end{subfigure}
  \hfill
  \begin{subfigure}[t]{0.24\textwidth}
    \includegraphics[width=\linewidth]{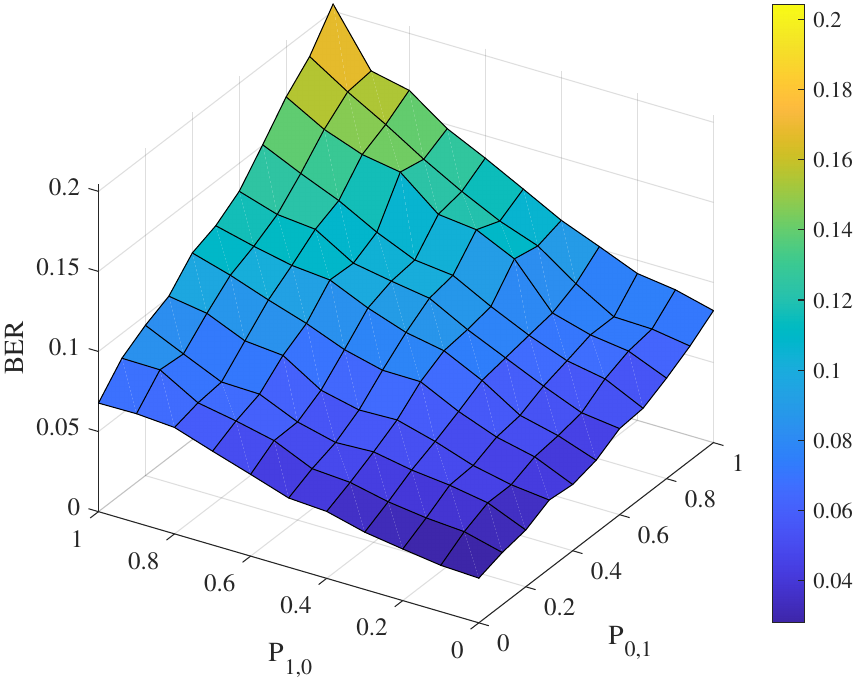}
    \captionsetup{skip=4pt}
    \caption{\footnotesize $N=5$}
    \label{fig:sub63}
  \end{subfigure}
  \begin{subfigure}[t]{0.24\textwidth}
    \includegraphics[width=\linewidth]{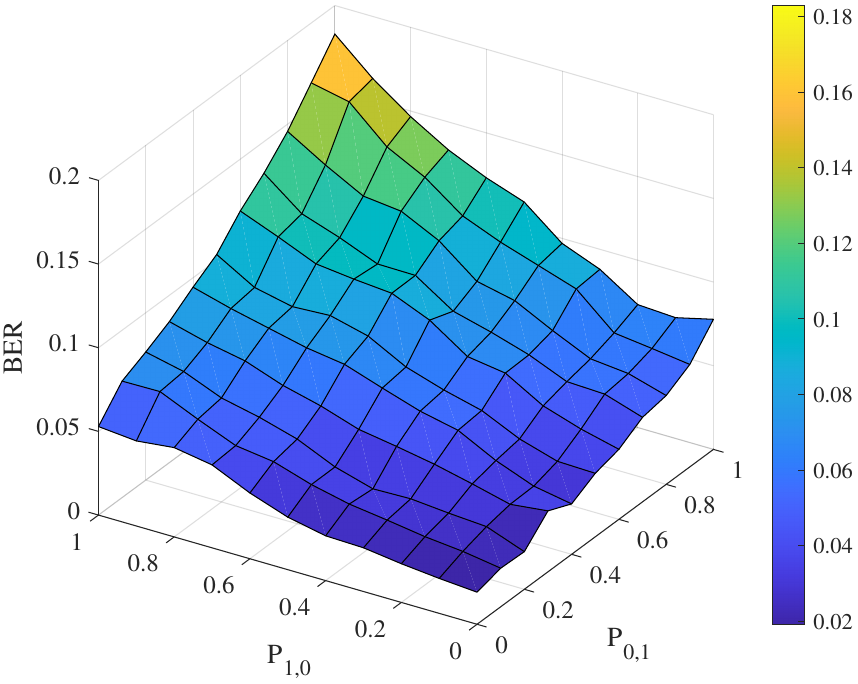}
    \captionsetup{skip=4pt}
    \caption{\footnotesize $N=7$ }
    \label{fig:sub64}
  \end{subfigure}
  \captionsetup{skip=4pt}
  \caption{The BER Performance of the Optimal Fusion Rule under Varying Numbers of RIS Reflecting Elements and SUs.}
  \label{fig6}
\end{figure*}

Nevertheless, even under favorable conditions, the AF attack remains the most destructive strategy, consistently causing the greatest degradation. Fig. \ref{fig6} shows that without attacks, more SUs reduce BER and improve performance, whereas under a fixed attack ratio, BER degradation stays stable across different SU counts, underscoring the robustness of this attack. Increasing RIS elements lowers BER through enhanced signal manipulation and reception, yet the AF attack still causes the largest performance drop.
In summary, across configurations involving attack ratios, SU antennas, SU numbers, and RIS elements, the AF attack consistently proves the most destructive and stable strategy, highlighting its critical role in system security analysis.

\begin{figure}[t]
  \centering
       \includegraphics[width=0.62\linewidth]{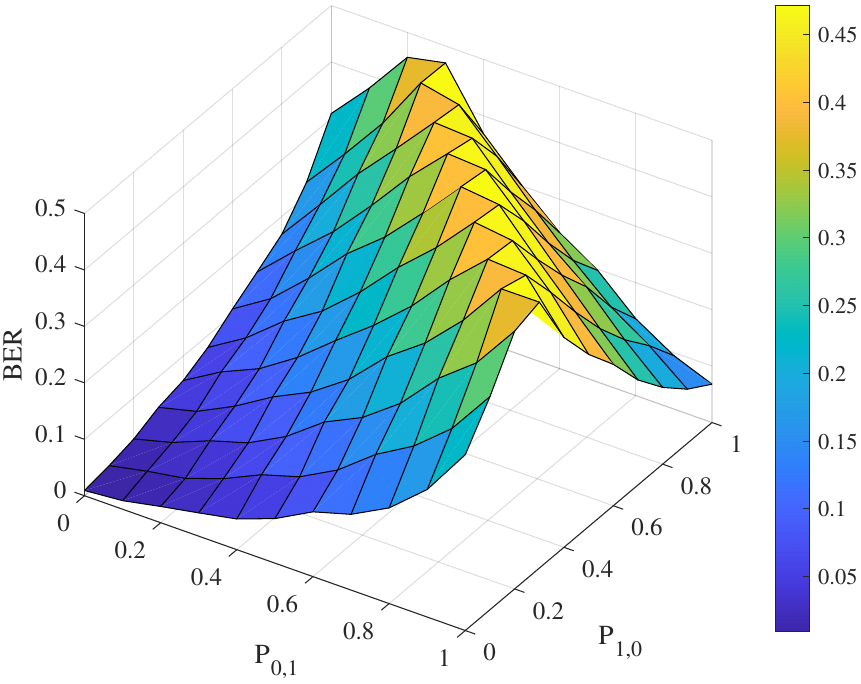}\\
       \captionsetup{skip=4pt}
  \caption{The BER Performance of Optimal Fusion Rule under Large-Scale Attacks.}\label{fig7}
\end{figure}

\begin{figure}[t]
  \centering
       \includegraphics[width=0.64\linewidth]{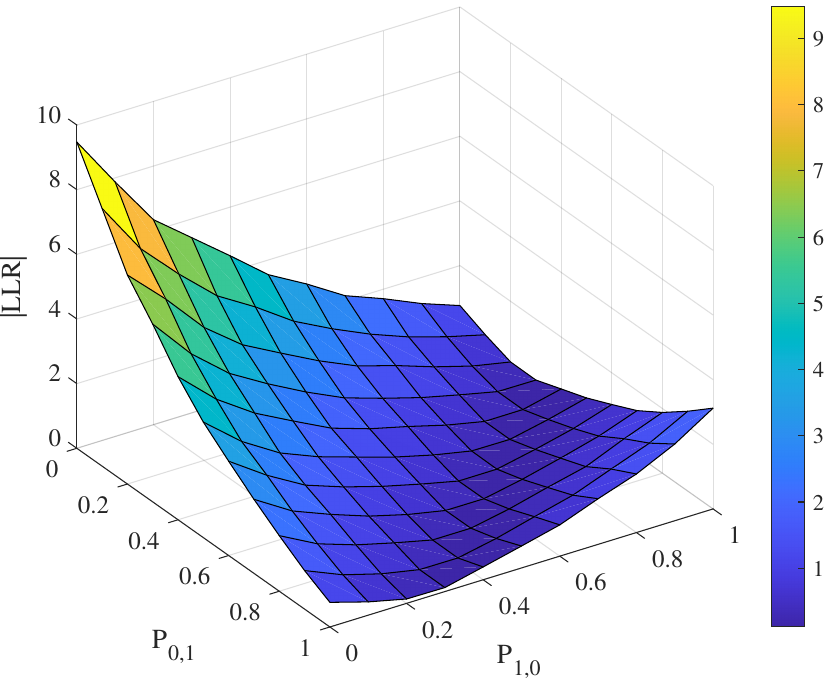}\\
       \captionsetup{skip=4pt}
  \caption{The Magnitude of LLR for Optimal Fusion Rules under Large-Scale Attacks.}\label{fig7-2}
\end{figure}

We first simulated the optimal large-scale Byzantine attack using default parameters to establish a baseline. As seen in Fig.~\ref{fig7}, detection accuracy drops sharply, and Fig.~\ref{fig7-2} shows the Magnitude of LLR collapsing to its minimum, signifying the attack's success in obscuring the true signal and undermining the FC's reliability.
Following this, we analyzed two key channel-related parameters to assess system robustness under fluctuations. Fig.~\ref{fig7} shows that under large-scale Byzantine attacks, the commonly used AF strategy ceases to be optimal. Specifically, when $\alpha(P_{0,1} + P_{1,0}) = 1$, the system's BER locks at 0.5, indicating complete corruption of the received signal and total loss of discriminability. In this state, no internal reconfiguration or adaptive defense can restore reliability, underscoring a fundamental system vulnerability to such powerful attacks.

\subsection{The Robustness Evaluation of Large-Scale Byzantine Attack Strategies under Varying Parameters}
\begin{figure*}[!t]
  \centering
  \begin{subfigure}[t]{0.24\textwidth}
    \includegraphics[width=\linewidth]{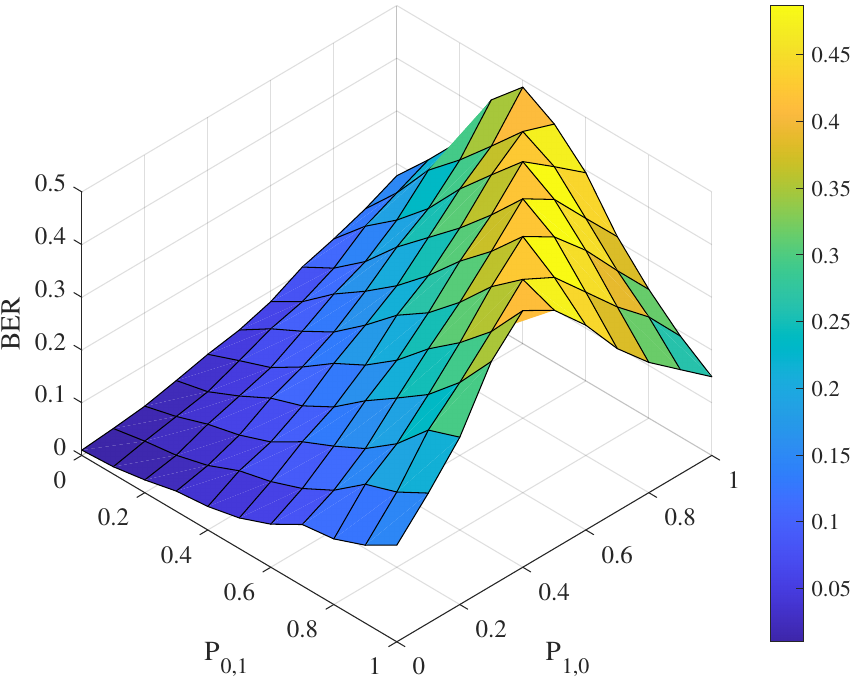}
    \captionsetup{skip=4pt}
    \caption{\footnotesize $\alpha=0.7$} 
    \label{fig:sub81}
  \end{subfigure}
  \hfill
  \begin{subfigure}[t]{0.24\textwidth}
    \includegraphics[width=\linewidth]{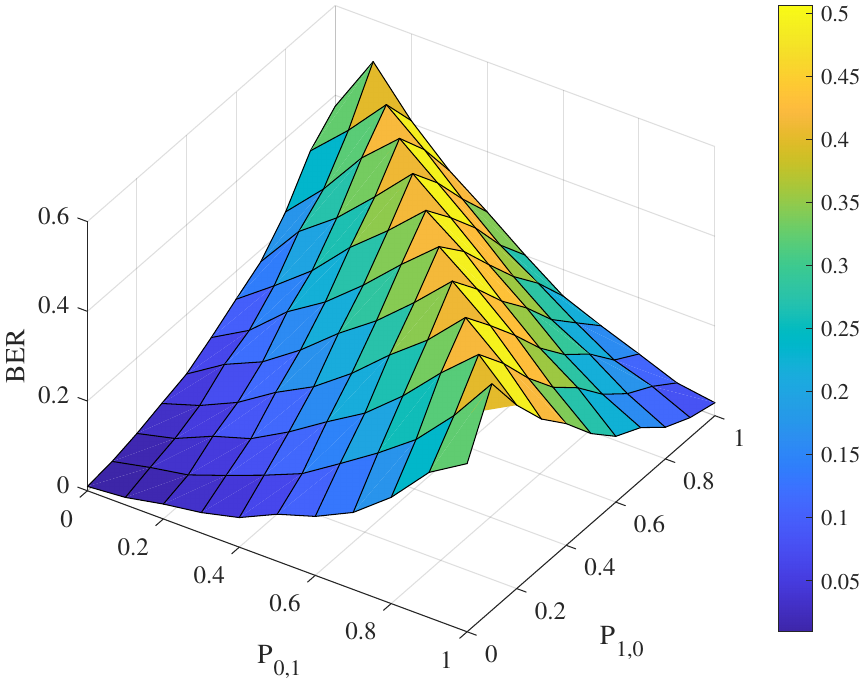}
    \captionsetup{skip=4pt}
    \caption{\footnotesize $\alpha=0.9$}
    \label{fig:sub82}
  \end{subfigure}
  \hfill
  \begin{subfigure}[t]{0.24\textwidth}
    \includegraphics[width=\linewidth]{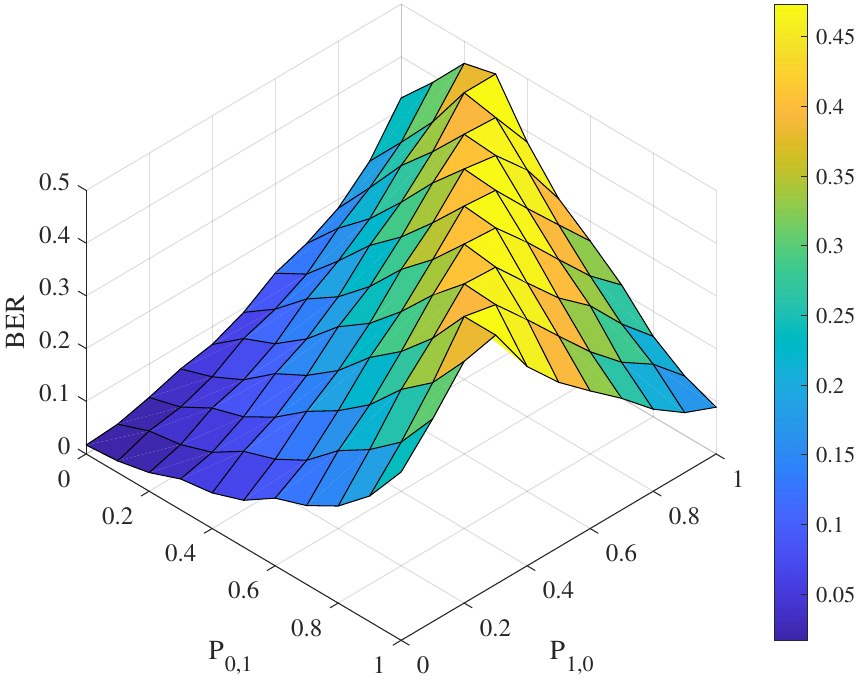}
    \captionsetup{skip=4pt}
    \caption{\footnotesize $M=4$}
    \label{fig:sub83}
  \end{subfigure}
  \begin{subfigure}[t]{0.24\textwidth}
    \includegraphics[width=\linewidth]{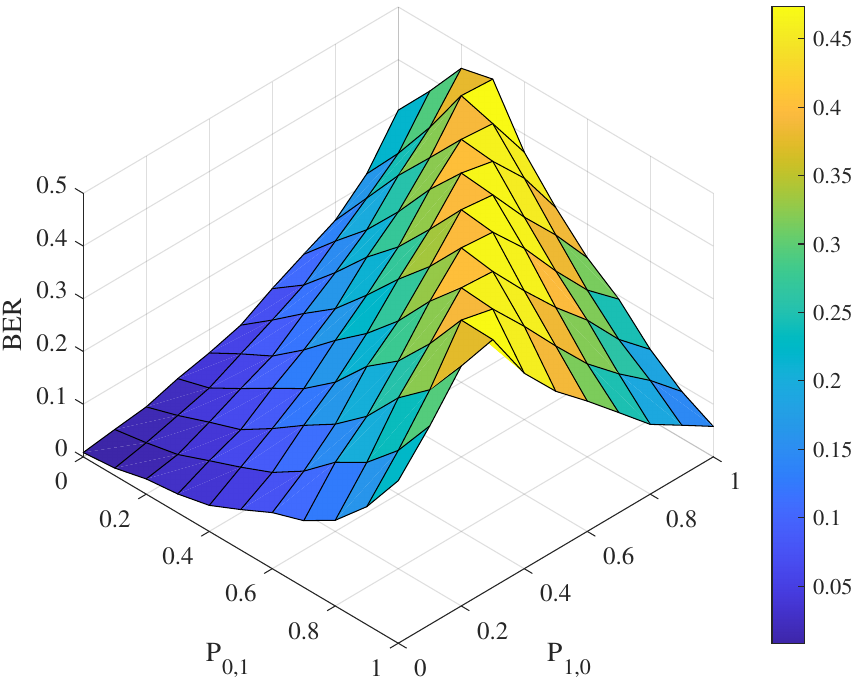}
    \captionsetup{skip=4pt}
    \caption{\footnotesize $M=8$}
    \label{fig:sub84}
  \end{subfigure}
  \caption{The BER performance of Optimal Fusion Rule under different Attack Strengths and Numbers of SUs Antennas.}
  \label{fig8}
\end{figure*}

When the attack ratio $\alpha$ increases from 0.7 to 0.9, the system exhibits significant performance changes. Specifically, while the baseline BER slightly decreases in the absence of attacks, indicating enhanced inherent anti-interference capability under attack conditions, the BER increases markedly. As shown in Fig. \ref{fig8}, when the attack parameters satisfy $\pi_{0,1} + \pi_{1,0} = 1$ (i.e., $\alpha \left( P_{0,1} + P_{1,0} \right) = 1$), the system reaches a worst-case attack scenario. Under this condition, the attacker can continuously force the BER to converge toward 0.5, effectively rendering the decision-making process unreliable. Notably, the destructive power of this attack strategy does not diminish with changes in $\alpha$. For example, when $\alpha = 0.9$, even partial flipping attacks can substantially degrade system performance. Compared to lower attack intensities, the BER is significantly higher, further demonstrating that the system becomes unrecoverable under large-scale attacks.

\begin{figure*}[!t]
  \centering
  \begin{subfigure}[t]{0.24\textwidth}
    \includegraphics[width=\linewidth]{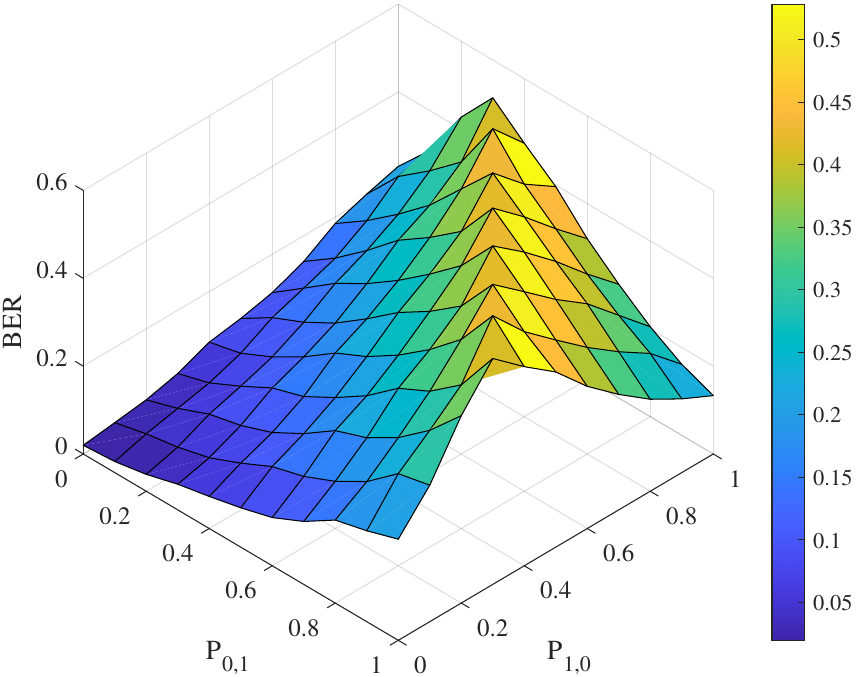}
    \captionsetup{skip=4pt}
    \caption{\footnotesize $I=8$} 
    \label{fig:sub91}
  \end{subfigure}
  \hfill
  \begin{subfigure}[t]{0.24\textwidth}
    \includegraphics[width=\linewidth]{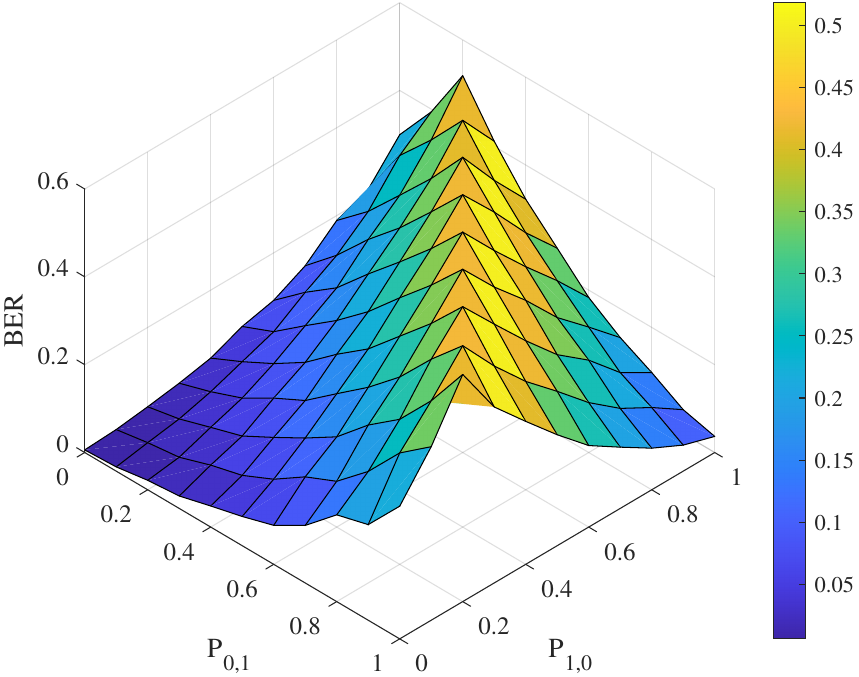}
    \captionsetup{skip=4pt}
    \caption{\footnotesize $I=12$}
    \label{fig:sub92}
  \end{subfigure}
  \hfill
  \begin{subfigure}[t]{0.24\textwidth}
    \includegraphics[width=\linewidth]{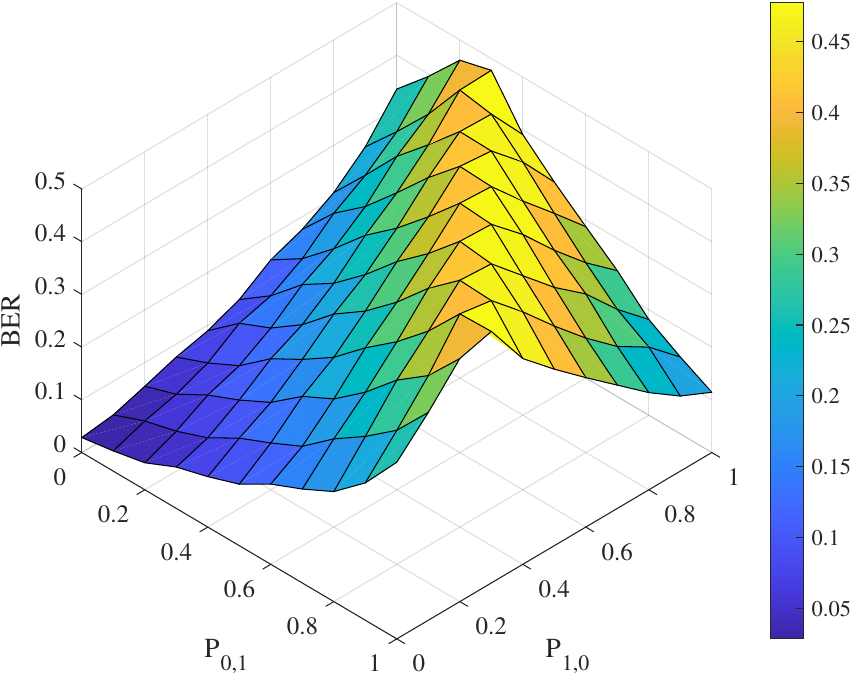}
    \captionsetup{skip=4pt}
    \caption{\footnotesize $N=5$}
    \label{fig:sub93}
  \end{subfigure}
  \begin{subfigure}[t]{0.24\textwidth}
    \includegraphics[width=\linewidth]{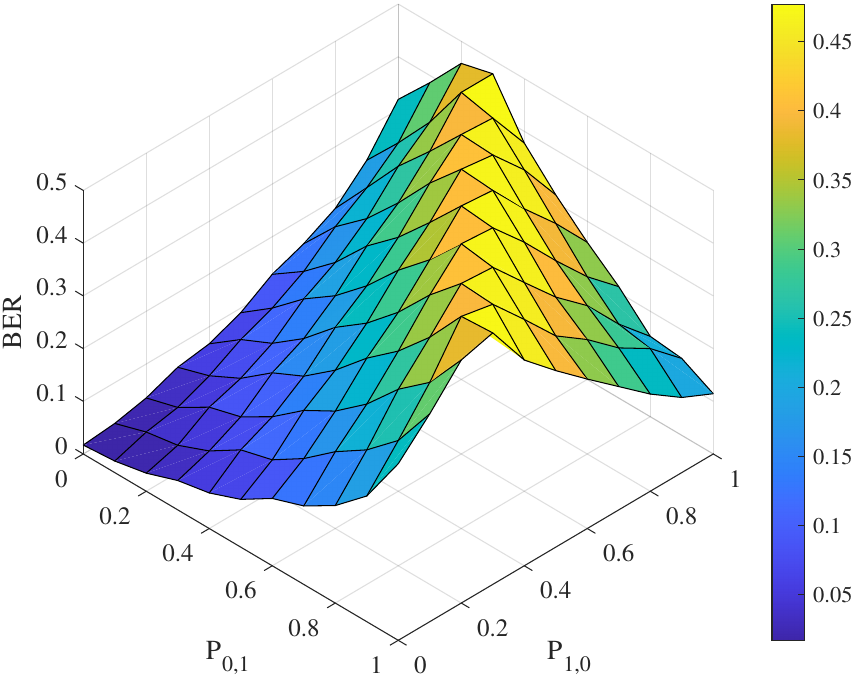}
    \captionsetup{skip=4pt}
    \caption{\footnotesize $N=7$}
    \label{fig:sub94}
  \end{subfigure}
  \captionsetup{skip=4pt}
  \caption{The BER Performance of the Optimal Fusion Rule under Varying Numbers of RIS Reflecting Elements and SUs.}
  \label{fig9}
\end{figure*}

With the attack ratio fixed at \(\alpha = 0.8\), the effect of varying the number of antennas \(M\) on system performance is studied. Without attacks, increasing antennas significantly improves BER due to spatial diversity gain. However, under optimal attacks, BER quickly converges to 0.5 for all antenna configurations, showing no mitigation of the attack's impact.
Similarly, with \(\alpha = 0.8\), increasing the number of SUs reduces BER without attacks, reflecting better reliability via cooperative detection. Yet, under optimal attacks, BER remains near 0.5 regardless of SU count, indicating attack resilience to user scale.
Fig.~\ref{fig9} shows BER versus transmission SNR for varying RIS elements \(N\). Increasing \(N\) greatly improves BER without attacks by enhancing diversity and noise suppression. Under optimal attacks, BER still approaches 0.5, suggesting attack effectiveness is largely unaffected by RIS size.
\subsection{The Strategy Analysis of Large-Scale Byzantine Attacks under the Optimal Fusion Rule}

In summary, regardless of system configurations, including attack ratio, antenna count, number of SUs, RIS scale, or relay node deployment, once the attack parameters satisfy $\alpha \left( P_{0,1} + P_{1,0} \right) = 1$, the system inevitably enters a decision failure state, with BER steadily converging to 0.5. This attack strategy demonstrates strong consistency and robustness across different network scales and attack intensities, making it one of the most destructive forms of Byzantine attacks.
\begin{figure}[t]
  \centering
       \includegraphics[width=0.65\linewidth]{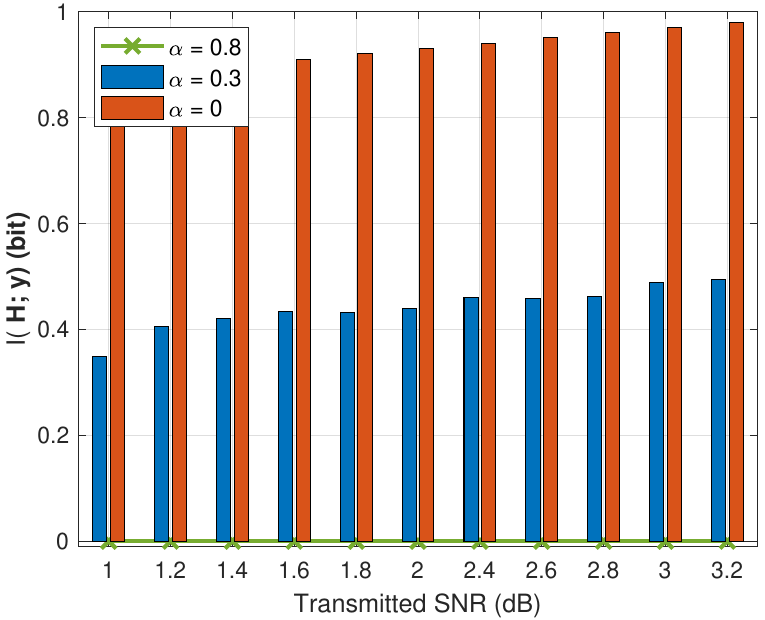}\\
       \captionsetup{skip=4pt}
  \caption{The MI Performance under Optimal Small-Scale and Large-Scale Attacks across Varying Transmitted SNRs.}\label{fig11}
\end{figure}

\subsection{Comparison of System Performance under Different Attack Modes}
A comparative analysis of system performance under varying adversarial conditions is conducted, with MI serving as the principal performance indicator. The study considers three scenarios: no attack, optimal small-scale attack, and optimal large-scale attack. As illustrated in Fig.~\ref{fig11}, the MI under optimal large-scale attack conditions remains nearly zero, signifying a substantial degradation in information exchange capability and overall system reliability. In contrast, although the optimal small-scale attack also leads to a decline in MI, its impact is markedly less severe. These observations demonstrate that the scale of adversarial intervention plays a decisive role in determining the extent of system disruption, underscoring the heightened risk posed by large-scale coordinated attacks.

\subsection{Evaluation of Different Attack Strategies for Small-Scale Byzantine Attacks}
\begin{figure*}[ht]
  \centering
  \begin{subfigure}[t]{0.24\textwidth}
    \includegraphics[width=\linewidth]{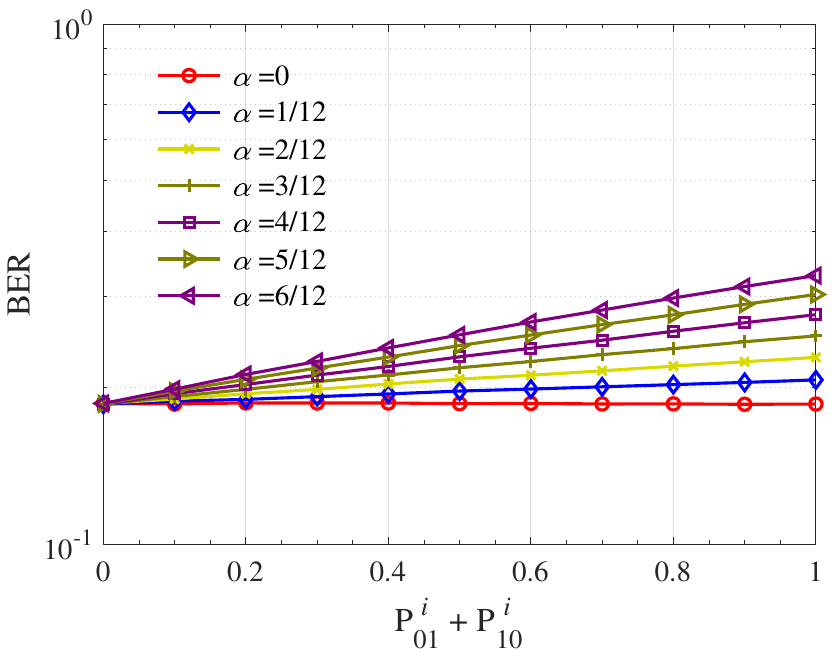}
    \captionsetup{skip=4pt}
    \caption{\footnotesize $0 \leqslant \alpha < \frac{1}{2}$, $0 \leqslant P_{01}^i + P_{10}^i \leqslant 1$} 
    \label{fig:sub101}
  \end{subfigure}
  \hfill
  \begin{subfigure}[t]{0.24\textwidth}
    \includegraphics[width=\linewidth]{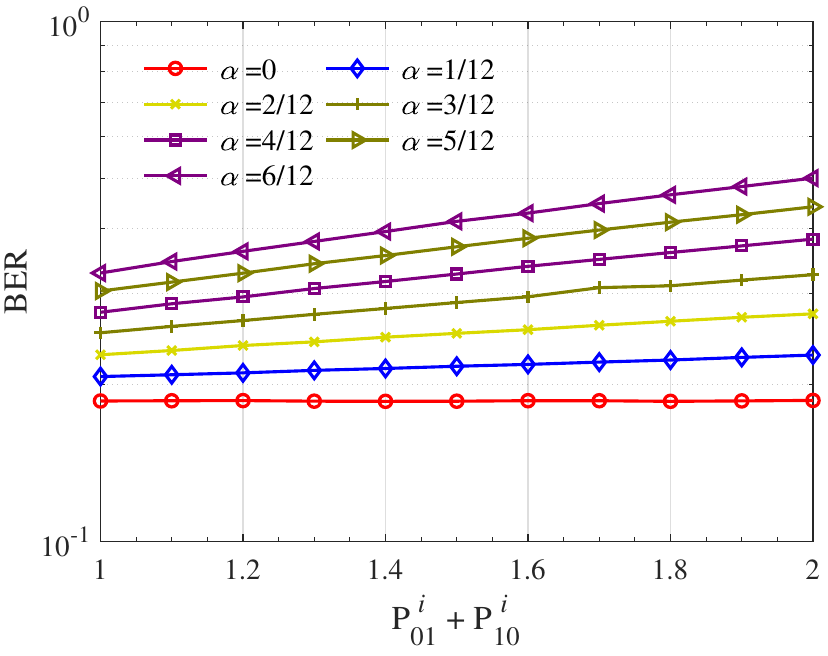}
    \captionsetup{skip=4pt}
    \caption{\footnotesize $0 \leqslant \alpha < \frac{1}{2}$, $1 \leqslant P_{01}^i + P_{10}^i \leqslant 2$}
    \label{fig:sub102}
  \end{subfigure}
  \hfill
  \begin{subfigure}[t]{0.24\textwidth}
    \includegraphics[width=\linewidth]{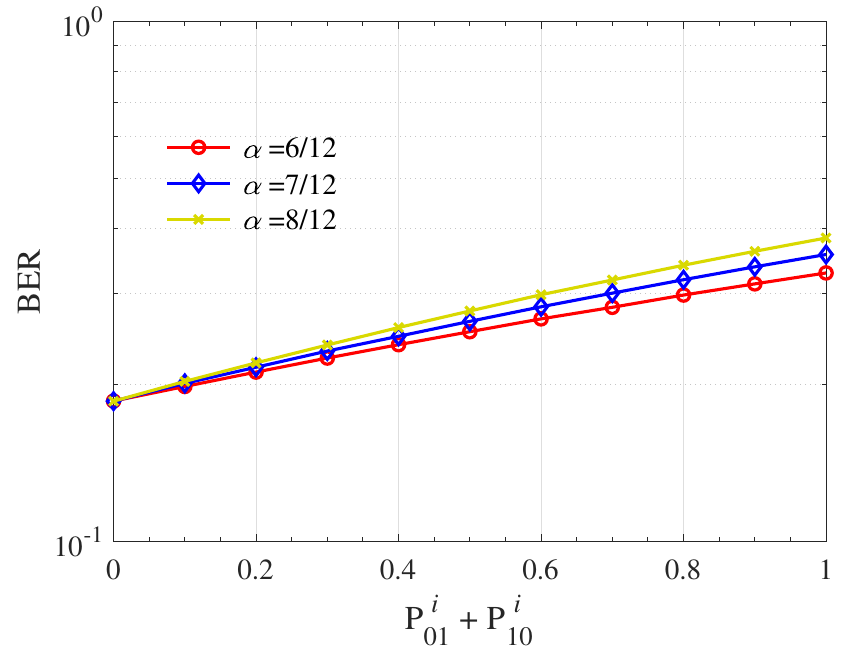}
    \captionsetup{skip=4pt}
    \caption{\footnotesize $\frac{1}{2} \leqslant \alpha \leqslant \frac{2}{3}$, $0 \leqslant P_{01}^i + P_{10}^i \leqslant 1$}
    \label{fig:sub103}
  \end{subfigure}
  \begin{subfigure}[t]{0.24\textwidth}
    \includegraphics[width=\linewidth]{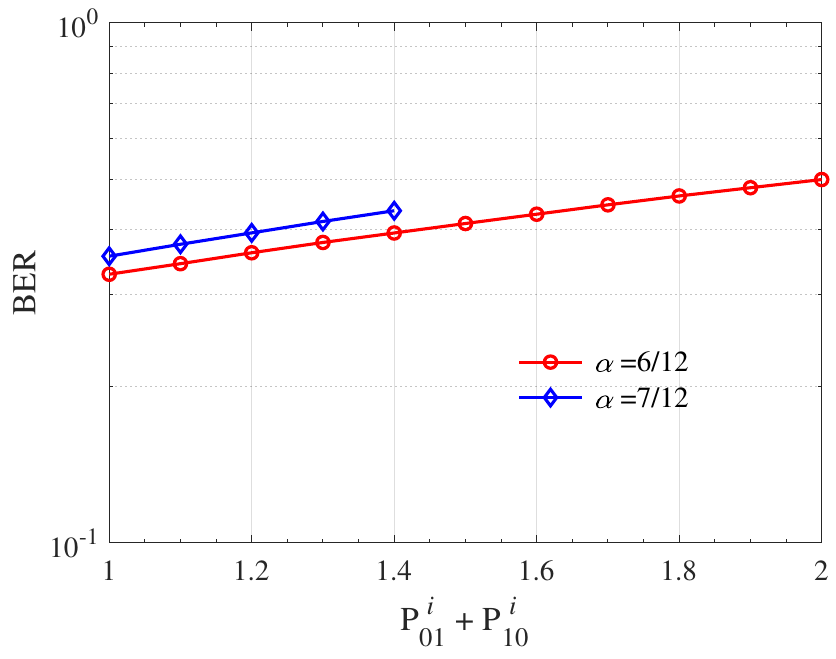}
    \captionsetup{skip=4pt}
    \caption{\footnotesize $\frac{1}{2} \leqslant \alpha \leqslant \frac{2}{3}$, $1 \leqslant {P_{01}^i + P_{10}^i} \leqslant {\frac{2}{\alpha } - 2}$ }
    \label{fig:sub104}
  \end{subfigure}
  \hfill
  \begin{subfigure}[t]{0.24\textwidth}
    \includegraphics[width=\linewidth]{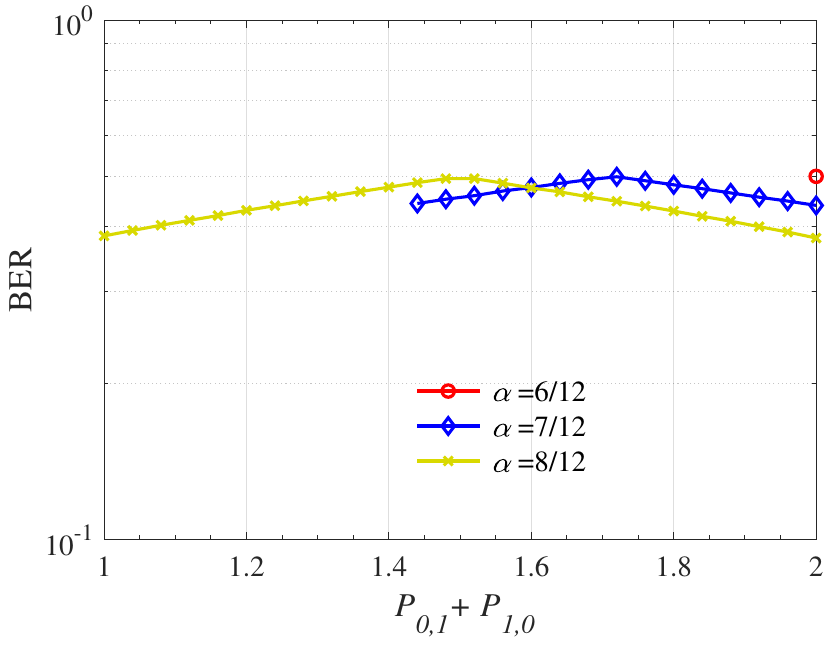}
    \captionsetup{skip=4pt}
    \caption{\footnotesize $\frac{1}{2} \leqslant \alpha \leqslant \frac{2}{3}$, ${\frac{2}{\alpha } - 2} \leqslant  {P_{01}^i + P_{10}^i}  \leqslant 2$}
    \label{fig:sub105}
  \end{subfigure}
  \hfill
  \begin{subfigure}[t]{0.24\textwidth}
    \includegraphics[width=\linewidth]{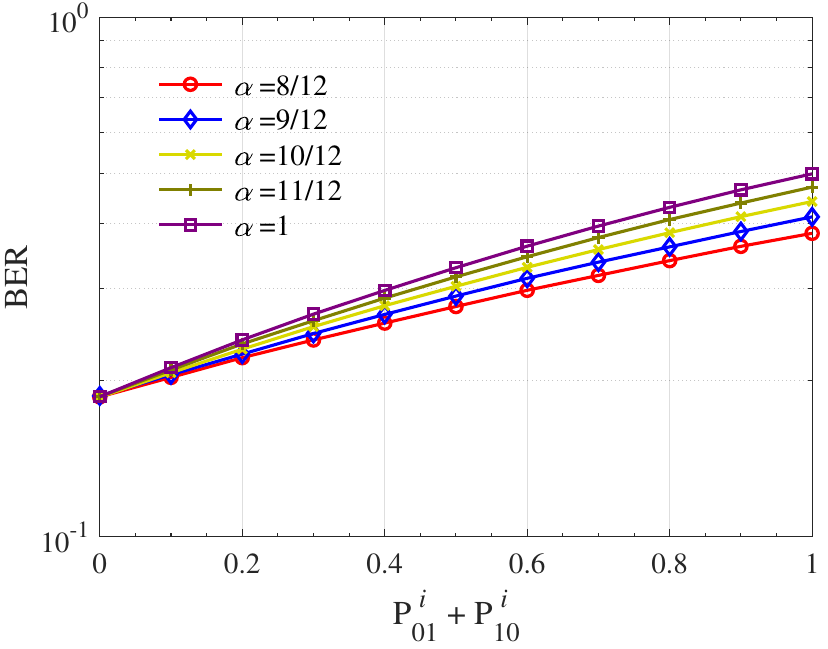}
    \captionsetup{skip=4pt}
    \caption{\footnotesize $\frac{2}{3} \leqslant \alpha \leqslant 1$, $0 \leqslant P_{01}^i + P_{10}^i \leqslant 1$}
    \label{fig:sub106}
  \end{subfigure}
  \begin{subfigure}[t]{0.24\textwidth}
    \includegraphics[width=\linewidth]{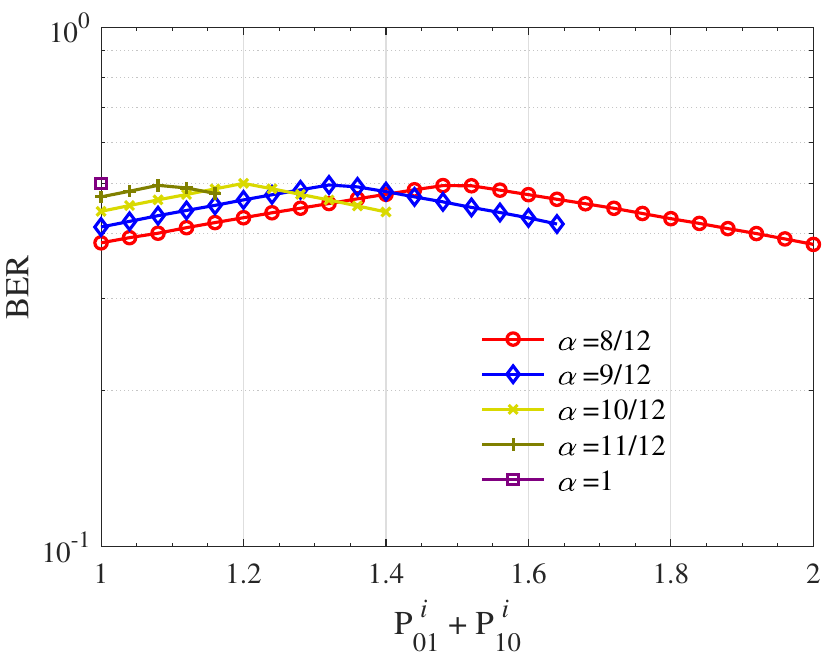}
    \captionsetup{skip=4pt}
    \caption{\footnotesize $\frac{2}{3} \leqslant \alpha \leqslant 1$, $1 \leqslant{P_{01}^i + P_{10}^i}\leqslant{\frac{2}{\alpha }-1}$}
    \label{fig:sub107}
  \end{subfigure}
  \hfill
  \begin{subfigure}[t]{0.24\textwidth}
    \includegraphics[width=\linewidth]{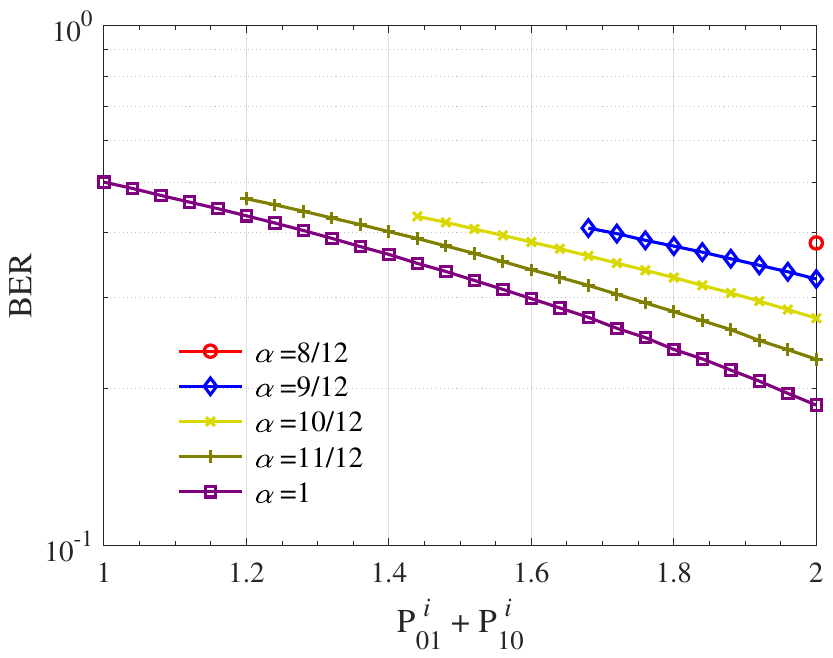}
    \captionsetup{skip=4pt}
    \caption{\footnotesize $\frac{2}{3} \leqslant \alpha \leqslant 1$, ${\frac{2}{\alpha } - 1} \leqslant {P_{01}^i + P_{10}^i} \leqslant 2$}
    \label{fig:sub108}
  \end{subfigure}
  \captionsetup{skip=4pt}
  \caption{Evaluating attack performance with Byzantine proportions}
  \label{fig12}
\end{figure*}
\noindent In this section, we evaluate the impact of different Byzantine attack strategies under small scale attacks. Fig. \ref{fig:sub101} and Fig. \ref{fig:sub102} illustrate Byzantine attacks within the range $0 \leq \alpha \leq \frac{1}{2}$.  Fig. \ref{fig:sub101} corresponds to attack scenarios where the flipping probabilities satisfy $0 \leq P_{0,1} + P_{1,0} \leq 1$. Fig. \ref{fig:sub102} presents cases where the flipping probabilities satisfy $1 \leq P_{0,1} + P_{1,0} \leq 2$.

From Fig. \ref{fig:sub101}, we observe that as the attack probability $\alpha$ increases, the system performance deteriorates. Under these conditions, both the AY and AN attacks have a greater impact than random attacks. In contrast, Fig. \ref{fig:sub102}, where the flipping probabilities satisfy $1 \leq P_{0,1} + P_{1,0} \leq 2$, allows for the possibility of a AF attack. The results indicate that the AF attack is the most detrimental, followed by RF attacks, while the AY and AN attacks have the least impact.
   
\subsection{Evaluation of Different Attack Strategies for Large Scale Byzantine Attacks}
\noindent In this section, we evaluate the impact of different Byzantine attack strategies under large scale attacks. Based on the previous analysis, the optimal attack strategy varies depending on the Byzantine node proportion $\alpha$. Therefore, we divide $\alpha$ into two intervals for analysis: $\frac{1}{2} \leq \alpha \leq \frac{2}{3}$ and $\frac{2}{3} \leq \alpha \leq 1$. In our experiments, we set the number of hops to $J$=3, the total number of sensors to $N$=12.  

Fig. \ref{fig:sub103} -- Fig. \ref{fig:sub105} illustrate the Byzantine attack scenarios when $\frac{1}{2} \leq \alpha \leq \frac{2}{3}$, while Fig. \ref{fig:sub106}-(f) -- Fig. \ref{fig:sub108} correspond to the cases where $\frac{2}{3} \leq \alpha \leq 1$.

Fig. \ref{fig:sub103} depicts the attack performance when the Byzantine ratio is within $\frac{1}{2} \leq \alpha \leq \frac{2}{3}$ and the flip probability satisfies $0 \leq P_{0,1} + P_{1,0} < 1$. We can observe that for a fixed number of attack nodes, the BER of AY and AN attacks is lower than that of RD attacks, indicating that AY and AN attacks are less disruptive than RD attacks in this scenario.  

Fig.~\ref{fig:sub104} examines attack performance for $\frac{1}{2} \leq \alpha \leq \frac{2}{3}$ in the probability range $1 \leq P_{0,1} + P_{1,0} \leq \frac{2}{\alpha}-2$. Under these conditions, the AF attack is most effective, followed by RD, while AY and AN show the weakest impact.

Fig.~\ref{fig:sub105} complements this analysis for the higher probability range, $\frac{2}{\alpha}-2 < P_{0,1} + P_{1,0} \leq 2$. Here, the performance ranking shifts: the RD attack becomes the most effective, outperforming the AF attack. The AY and AN attacks remain the least impactful. The case for $\alpha=\frac{1}{2}$ is excluded, as its corresponding range is trivial.


Fig. \ref{fig:sub106} presents the attack performance under the conditions where the Byzantine ratio satisfies \( \frac{2}{3} \leq \alpha \leq 1 \) and the flip probability is within the range $0 \leq P_{0,1} + P_{1,0} < 1$. From the figure, it can be observed that when the number of attacking nodes is fixed, the bit error rate of AY and AN attacks is higher than that of the RD attack, indicating that the AY and AN attacks are stronger than the RD attack under these conditions.

Fig.~\ref{fig:sub107} investigates attack performance for Byzantine ratios $\frac{2}{3} \leq \alpha < 1$ within the flip probability range of $1 \leq P_{0,1} + P_{1,0} \leq \frac{2}{\alpha}-1$. Results show that for a fixed number of attacking nodes, the RD attack is most effective, followed by AY and AN, while AF is least effective. The case $\alpha=1$ is omitted as its valid probability range collapses to a point.

Fig.~\ref{fig:sub108} extends the analysis to the higher flip probability range of $\frac{2}{\alpha}-1 < P_{0,1} + P_{1,0} \leq 2$. In this regime, the attack performance ranking changes: the AY and AN attacks become the most effective, surpassing the RD attack. The AF attack remains the weakest.

\section{CONCLUSIONS AND FUTURE WORK}\label{section10}
We have studied the problem of optimal Byzantine attack strategies in CSS under non-ideal conditions. Assuming that the attackers are aware of the fusion rule, we propose optimal attack strategies for both optimal and suboptimal fusion rules. Our findings indicate that even when Byzantine attackers are unaware of the current fusion strategy, they can still launch optimal attacks. Furthermore, we evaluate the performance of different attack strategies under various conditions. Our future research directions will be in the following areas.
We plan to extend the existing optimal attack strategies from the traditional binary hypothesis testing framework to more complex multi-channel spectrum sensing scenarios \cite{mughal2024intelligent,10866626,10716953}. In this extended setting, our focus will be on investigating optimal attack behaviors in multi-dimensional signal spaces \cite{10225318}, particularly how attackers can formulate coordinated interference strategies under varying channel conditions and the coexistence of multiple signal features to maximize disruption of the decision-making process \cite{10666007,10632743,10506656}.
In addition, We will further explore defenses against optimal or near-optimal attacks. Based on existing results, lightweight and effective detection and defense schemes will be designed to exploit statistical inconsistencies caused by attackers, improving the robustness and recovery of CSS systems under adversarial conditions.

\section*{Appendix A}\label{Appendix A}
\subsubsection{AN attacks and AY attacks}
The numerator term in an AN attack is roughly proportional to $-{{P}_{{D}_{i}}}\left( 1-\alpha \right)$. The denominator, on the other hand, is proportional to $-{{P}_{{F}_{i}}}\left( 1-\alpha \right)$, so the overall decision statistic should be proportional to $-\left( {{P}_{{D}_{i}}}-{{P}_{{F}_{i}}} \right)\left( 1-\alpha\right)$ is proportional.

Similarly the numerator is roughly proportional to 
$-\left[ \alpha +\left( 1-\alpha \right){P}_{{{D}_{i}}} \right]$ for an AY attack. 
The denominator, on the other hand, is proportional to $-\left[ \alpha +\left( 1-\alpha \right){P}_{{{F}_{i}}} \right]$, 
so the overall decision statistic should be proportional to $-\left( {{P}_{{D}_{i}}}-{{P}_{{F}_{i}}} \right)\left( 1-\alpha \right)$ is proportional.

That is, both AN attacks and AY attacks have their overall decision statistics proportional to $-\left( {{P}_{{{D}_{i}}}}-{{P}_{{F}_{i}}} \right)\left( 1-\alpha \right)$. This leads to the fact that the attack performance of AN and AY attacks is considered to be the same regardless of whether it is a large-scale attack or a small-scale attack.

\subsubsection{AF Attack and AN Attack}
According to (\ref{math54}), the numerator in the AF attack is approximately proportional to: $-\left[ \alpha -\left( 2\alpha -1 \right) P_{D_i} \right]$. The denominator is approximately proportional to:$-\left[ \alpha -\left( 2\alpha -1 \right) P_{F_i} \right]$.
Therefore, the overall decision statistic is proportional to: $\left( P_{D_i} - P_{F_i} \right) \left( 2\alpha -1 \right)$. Since it is known that \( P_{D_i} - P_{F_i} > 0 \), comparing \( (1 - \alpha) \) and \( |2\alpha - 1| \) can help determine which attack results in a smaller magnitude of the decision statistic.

\begin{itemize}
  \item \textbf{Small-scale attack}: When \( \alpha < 0.5 \), it can be concluded that under small-scale attacks, the AF attack outperforms the AN attack.
  \item \textbf{Large-scale attack}: when \( \frac{1}{2} \le \alpha \le \frac{2}{3} \), the AF attack performs better. When \( \frac{2}{3} \le \alpha \le 1 \), the AN or AY attacks perform better.
\end{itemize}

\subsubsection{RF Attacks and AN Attacks} 
For the RF attack, the numerator is approximately proportional to:

$-\left[ P_{D_i} + \alpha P_{1,0} - P_{D_i} \alpha (P_{0,1} + P_{1,0}) \right]$. 

The denominator is approximately proportional to: 

$-\left[ P_{F_i} + \alpha P_{1,0} - P_{F_i} \alpha (P_{0,1} + P_{1,0}) \right]$. 

Therefore, the overall decision statistic is proportional to: 

$-\left( P_{D_i} - P_{F_i} \right) \left[ \alpha (P_{0,1} + P_{1,0}) - 1 \right]$.

To compare the performance of RF and AN attacks, we use $\left| \alpha (P_{0,1} + P_{1,0}) - 1 \right|$  and $(1-\alpha)$ to determine which yields a smaller magnitude of the decision statistic

\begin{itemize}
  \item \textbf{Small-scale attack}: since \( \alpha < 0.5 \), by comparing \( 1 - \alpha (P_{0,1} + P_{1,0}) \) and \( (1 - \alpha) \), we can conclude that: when \( 0 \le P_{0,1} + P_{1,0} \le 1 \), AN attacks perform better. When \( 1 \le P_{0,1} + P_{1,0} \le 2 \), RF attacks perform better.
  \item \textbf{Large-scale attack}: since $\alpha > 0.5$, we have 
  
  $\left| \alpha (P_{0,1} + P_{1,0}) - 1 \right| = \alpha (P_{0,1} + P_{1,0}) - 1$. 

  When \( 0 \le P_{0,1} + P_{1,0} \le 1 \),  it can be concluded that \( (1 - \alpha) \ge \left| \alpha (P_{0,1} + P_{1,0}) - 1 \right| \), meaning AN or AY attacks always perform better than RF attacks. When \( 1 \le P_{0,1} + P_{1,0} \le 2 \), by setting \( (1 - \alpha) = \alpha (P_{0,1} + P_{1,0}) - 1 \), we can find that \( P_{0,1} + P_{1,0} = \frac{2}{\alpha} - 1 \).  When \( 1 \le P_{0,1} + P_{1,0} \le \frac{2}{\alpha} - 1 \), RF attacks perform better. When \( \frac{2}{\alpha} - 1 \le P_{0,1} + P_{1,0} \le 2 \), AN attacks perform better.
\end{itemize}

\subsubsection{RF Attacks and AF Attacks}
Compare the performance of RF attacks and AF attacks, we can use \( \left| \alpha (P_{0,1} + P_{1,0}) - 1 \right| \) and \( \left( 2\alpha -1 \right) \) to determine which attack performs better.

Following similar analysis method, its easy to conclude that 
\begin{itemize}
  \item \textbf{Small-scale attack}: When $0 \le P_{0,1} + P_{1,0} \le 1$, AF attack is impossible because they require $P_{0,1} = P_{1,0} = 1$. When 1 $\le P_{0,1} + P_{1,0} \le 2$, AF attacks perform better.

\item \textbf{Large-scale attack}: when \( 1 \le P_{0,1} + P_{1,0} \le \frac{2}{\alpha} - 2 \), AF attacks perform better. When \( \frac{2}{\alpha} - 2 \le P_{0,1} + P_{1,0} \le 2 \), RF attacks perform better.
\end{itemize}

\bibliographystyle{IEEEtran}
\bibliography{mybibfile.bib}

\end{document}